\documentclass[prd,showkeys,preprintnumbers,floatfix,
nofootinbib,superscriptaddress]{revtex4-1}
\usepackage{float}
\usepackage{slashed}
\usepackage{mathtools}
\usepackage{amsfonts} % AMS
\usepackage{amssymb} % AMS
\usepackage{amsmath} % AMS
\usepackage{graphicx} % Include figure files
\usepackage{array} % array
\usepackage{dcolumn} % Align table columns on decimal point
\usepackage{bm} % bold math
\usepackage{latexsym} % latex symbols
\usepackage{longtable} % long tables
\usepackage{hyperref} % hypertext links 
\usepackage{verbatim}
\usepackage{epsfig}
\usepackage{color}
\newcommand{\Mthr}[0]{\mathcal M_{3,\thr}}

\newcommand{\thr}[0]{\textrm{thr}}
\newcommand{\disc}[0]{\textrm{disc}}
\newcommand{\conn}[0]{\textrm{conn}}
\newcommand{\CZ}[2]{C_{#1,\thr}^{(#2)}(0)}
\newcommand{\CZcon}[2]{C_{#1,\thr,\conn}^{(#2)}(0)}
\newcommand{\CT}[2]{\partial_\tau C^{(#2)}_{#1,\thr}(0)}
\newcommand{\CTcon}[2]{\partial_\tau C^{(#2)}_{#1,\thr,\conn}}
\newcommand{\CZZ}[1]{C_{#1,\thr}(0)}
\newcommand{\CTT}[1]{\partial_\tau C_{#1,\thr}(0)}

\newcommand{\off}[0]{\textrm{off}}

\newcommand{\ct}[0]{\textrm{ct}}
\newcommand{\bare}[0]{\textrm{bare}}

\newcommand{\cI}[0]{\mathcal I}
\newcommand{\cJ}[0]{\mathcal J}
\newcommand{\cO}[0]{\mathcal O}
\newcommand{\cM}[0]{\mathcal M}

\newcommand{\cK}[0]{\mathcal K}

\newcommand{\cS}[0]{\mathcal S}
\newcommand{\cC}[0]{\mathcal C}

\newcommand{\four}[0]{{(4)}}
\newcommand{\US}[0]{{\rm US}}
\newcommand{\ST}[0]{{\rm ST}}
\newcommand{\SU}[0]{{\rm SU}}

\newcommand{\HSPT}[0]{\cite{Hansen:2015zta}}
\newcommand{\HSTH}[0]{\cite{Hansen:2016fzj}}

\newcommand{\HSQCb}[0]{\cite{Hansen:2015zga}}
\newcommand{\HSQCa}[0]{\cite{Hansen:2014eka}}
\newcommand{\PS}[0]{\cite{Peskin:1995ev}}
\newcommand{\Beane}[0]{\cite{Beane2007}}
\begin{document}
% \preprint{\vbox{\hbox{JLAB-THY-17-2400} }}
\title{Testing the threshold expansion for three-particle energies at fourth order
in $\phi^4$ theory}
\author{Stephen R. Sharpe}
\email[e-mail: ]{srsharpe@uw.edu}
\affiliation{
Physics Department, University of Washington, Seattle, WA 98195-1560, USA\\
}

\date{\today}
\begin{abstract}
A relativistic formalism for relating 
the energies of the states of three scalar particles in finite volume 
to infinite volume scattering amplitudes has recently been developed.
This formalism has been used to predict the
energy of the state closest to threshold in an expansion in powers of $1/L$,
with $L$ the box length.
This expansion has been tested previously by a perturbative calculation of the threshold energy
in $\lambda \phi^4$ theory, working to third order in $\lambda$ 
and up to $\cO(1/L^6)$ in the volume expansion.
However, several aspects of the predicted threshold behavior
do not enter until fourth (three-loop) order in perturbation theory.
Here I extend the perturbative calculation to fourth order and find agreement
with the general prediction.
This check also requires a two-loop calculation of the 
infinite-volume off-shell two-particle scattering amplitude near threshold.
As a spin-off,  I check the threshold expansion for two particles to the same order,
finding agreement with the result that follows from L\"uscher's formalism.

\end{abstract}

\keywords{finite volume, relativistic scattering theory, lattice QCD}
\maketitle

\section{Introduction}

There is considerable  interest in developing 
theoretical formalism to allow lattice QCD to
determine the properties of resonances 
for which some of the decay channels involve
three or more particles. Such formalism is needed for the study of most of 
the strong-interaction resonances that appear in nature, 
e.g.~the $\omega$ meson and the Roper baryon.
Specifically, what is needed on the theoretical side is a quantization condition
that relates the energies of multiparticle states in a finite volume to the infinite-volume
scattering amplitudes of these particles.
While such a quantization has long been known for two particles
(based on Refs.~\cite{Luscher:1986n2,Luscher:1991n1} and subsequent generalizations),
the three particle quantization condition is relatively 
new~\cite{Hansen:2014eka,Hansen:2015zga,Briceno:2017tce}
(and not yet completely general).
Since the formalism is rather involved, it is important to provide detailed checks
that test all aspects of the approach.\footnote{%
An alternative approach to the three-particle
quantization condition was proposed very recently in Refs.~\cite{Hammer:2017uqm,Hammer:2017kms}.
The threshold expansion in this new approach has not yet been derived.}

The present work is aimed at extending previous tests 
of the formalism of Refs.~\cite{Hansen:2014eka,Hansen:2015zga,Briceno:2017tce}
by considering the prediction of
the quantization condition for a system of three identical scalar particles near threshold.
These particles are confined to a cubic box of side $L$ (as in a lattice simulation)
and it is assumed that there is a $Z_2$ symmetry restricting interactions to those
involving an even number of particles. The total momentum\footnote{%
I use "momentum" for three-momentum throughout this work aside from in
Appendix~\ref{app:K}.}
is taken to be zero. Under these assumptions, Ref.~\HSTH\
derived the expansion of the energy of the three-particle threshold state
in powers of $1/L$, keeping terms up to $\cO(1/L^6)$.
This threshold expansion was derived for an arbitrary $Z_2$-symmetric
effective field theory.
Unlike the two-particle case, where the derivation
of the threshold expansion is rather straightforward, the derivation for three particles
is itself very involved, requiring the summation of several infinite series.
Thus the test presented here is a check of the derivation of the threshold expansion
as well as of the underlying formalism.

The general formula for the threshold expansion is given in terms of infinite-volume
quantities such as the two-particle scattering length. This result is tested here
by calculating the same expansion in 
a specific $Z_2$-symmetric theory---$\lambda \phi^4$ theory---and expressing the result
in terms of the same infinite-volume quantities.
This test has previously been passed  at third order in $\lambda$,
and through $\cO(1/L^6)$ in the volume expansion, in Ref.~\HSPT, 
and what is presented here is the fourth-order calculation to the same order in $1/L$.
The specific motivation for carrying out this lengthy and quite tedious calculation
is that the fourth-order calculation tests qualitatively new aspects of the general prediction.
Specifically, the general formalism contains a ``divergence-free" three-particle
scattering amplitude that is obtained from the three-to-three amplitude $\cM_3$
by  subtracting an infinite series of terms such that the physical singularities are removed.
I stress that such singularities are inevitably present in $\cM_3$ and must be dealt with.
A simplified version of this subtraction 
procedure is sufficient at threshold~\HSTH, and defines a quantity called $\Mthr$.
The $\cO(\lambda^3)$ calculation did not test all the subtraction terms in the
definition of $\Mthr$, but the present calculation does.

It turns out that, as part of the calculation of the three-particle threshold energy,
one needs all the ingredients necessary to determine the two-particle threshold energy.
Thus the latter energy can also be compared to the general result that follows from the
formalism of Refs.~\cite{Luscher:1986n2,Luscher:1991n1}. 
Since by now there is no doubt that this formalism is correct, this subsidiary
calculation provides a check on the methods used here.

This paper is organized as follows. 
The following section contains a summary of the methods introduced in Ref.~\HSPT\
to determine the threshold energy in perturbation theory, and presents the
general results from Ref.~\HSTH\ that are being tested.
Section~\ref{sec:DE24} concerns the two-particle energy shift, and provides a sketch
of the calculation and the final results. These require the two-loop contribution to the
effective range.
Section~\ref{sec:DE34} describes the calculation of the contributions to the
energy shift that are specific to three particles. This requires a particular off-shell
version of the two-loop infinite-volume scattering amplitude, the calculation of which
is similar to, but different from, that of the effective range.
I conclude in Sec.~\ref{sec:conc}.
Technical details are collected in three appendices: the first recalling some general
results for finite-volume sums, the second listing the needed counterterms, and the
third describing the calculation of the on- and off-shell two-loop scattering amplitude
near threshold.

\section{Overview of methods and results to be tested}
\label{sec:methods}

The method I use is that introduced in Ref.~\cite{Hansen:2015zta},
and I recall here only the essential features.
The theory has the Euclidean Lagrangian density 
\begin{equation}
{\cal L} =  \frac12 \partial_\mu\phi\partial_\mu \phi
+ \frac{m^2}2 \phi^2 + \frac{\lambda}{4!} \phi^4 
+ \frac{\delta Z}2   \partial_\mu\phi\partial_\mu \phi 
+ \frac{\delta Z_m}2  m^2 \phi^2
+  \frac{{\delta Z_\lambda}}{4!} \phi^4
\,,
\label{eq:action}
\end{equation}
with $\phi$ a scalar field. An on-shell renormalization scheme is used:
$\delta Z$ and $\delta Z_m$ are tuned
so that $m$ is the physical mass and the residue 
of the (infinite-volume) propagator at the pole is unity.
The counterterm $\delta Z_\lambda$ is defined by the requirement that the scattering
amplitude at threshold is given by $-\lambda$ to all orders.
Since this threshold amplitude is, by definition, proportional to the scattering length, $a$,
this renormalization condition implies the exact relation
\begin{equation}
\lambda = 32 \pi m a\,.
\label{eq:lambdadef}
\end{equation}
I will need the two-loop form of $\delta Z_\lambda$, and this is
given in Appendix~\ref{app:counter}.

The finite-volume (FV) energies are extracted from the long-time behavior of the
following correlation functions:
\begin{align}
\label{eq:C2def}
C_2(\tau) &=\frac{(2m)^2}{2 L^6} e^{2m\tau}
\left\langle \tilde \phi_{\vec 0}(\tau)^2
\tilde \phi_{\vec 0}(0)^2 \right\rangle \,,
\\
\label{eq:C3def}
C_3(\tau) &=\frac{(2m)^3}{6 L^9} e^{3m\tau}
\left\langle \tilde \phi_{\vec 0}(\tau)^3
\tilde \phi_{\vec 0}(0)^3 \right\rangle \,.
\end{align}
Here $\tau$ is Euclidean time, which is always taken to be positive or zero, 
and the % momentum-space 
interpolating fields are
\begin{equation}
\tilde \phi_{\vec p}(\tau) = \int_L d^3 x \; e^{-i \vec p \cdot \vec x} \phi(\vec x, \tau)\,,
\end{equation}
with the subscript $L$ indicating that the integral is over the cubic box.
Periodic boundary conditions are applied to $\phi$,
so that momenta are quantized as $\vec p = 2\pi \vec n/L$, with $\vec n$
a vector of integers.
Euclidean time is taken to have infinite range.
The prefactors in Eqs.~(\ref{eq:C2def}) and (\ref{eq:C3def}) are chosen so that
$C_j(\tau) =1$ for all $\tau$ if $\lambda=0$. In this limit the interpolating operators
couple to the states consisting of $j$ particles at rest.

When $\lambda \ne 0$ the correlators behave as  (recalling that $\tau \ge 0$)
\begin{equation}
C_j(\tau) = \sum_k  A_{j,k} \, e^{-\Delta E_{j,k} \tau}\,,
\qquad
\Delta E_{j,k} \equiv E_{j,k}-j m \,,
\label{eq:amps}
\end{equation}
where $j=2$ or $3$, $k$ labels the finite-volume states that couple to the interpolators, 
and $A_{j,k}$ are the corresponding amplitudes. 
The state of interest is that nearest threshold
for which $\Delta E_{j,k}\to 0$ as $\lambda \to 0$. This is labeled by $k=\thr$.
The procedure developed in Ref.~\cite{Hansen:2015zta} for picking out its energy
is to first calculate $C_2(\tau)$ and $C_3(\tau)$ order by order in perturbation theory,
then remove by hand exponentially growing or falling contributions.
The resulting subtracted correlators have the form 
\begin{equation}
C_{j,\thr}(\tau) = \CZZ j + \tau \big [ \CTT j \big ] + {\cal O}(\tau^2)\,.
\label{eq:Cjexpand}
\end{equation}
Finally, the shift of the desired energy from threshold is given by
\begin{equation}
\Delta E_{j,\thr} %= E_{j,\thr}-j m 
= - \frac{\CTT j}{\CZZ j}\,.
\label{eq:DeltaEmethod}
\end{equation}
The justification for this method is explained in Ref.~\HSPT.

The perturbative expansions of the quantities appearing in this expression 
are\footnote{%
Here I am expanding in the renormalized coupling,
whereas the corresponding expansions in Ref.~\cite{Hansen:2015zta}
were in powers of the bare coupling. For the sake of brevity,
I continue to use the same
notation for the expansion coefficients, although the values of these
coefficients differ.}
\begin{align}
\CZZ j &= 1 + \sum_{n=1}^\infty  \lambda^n \CZ j n \,,
\label{eq:Cjthr0exp}
\\
\CTT j &= \sum_{n=1}^\infty  \lambda^n \big  [\CT j n \big ] \,,
\label{eq:Cjthr1exp}
\\
\Delta E_{j,\thr} &= \sum_{n=1}^\infty  \lambda^n \Delta E_{j,\thr}^{(n)} \,.
\label{eq:DeltaEexp}
\end{align}
Inserting these expansions into Eq.~(\ref{eq:DeltaEmethod}), 
the result for the fourth order term in $\Delta E_{j,\thr}$ is
\begin{multline}
\Delta E_{j,\thr}^{(4)} = - \CT j 4 + \CZ j 1 \big [\CT j 3 \big ]
+ \CZ j 2 \big [\CT j 2 \big ] + \CZ j 3 \big [\CT j 1 \big ]\\
 - \big[\CZ j 1\big]^2 \big [\CT j 2\big ] 
 -  2 \CZ j 1 \CZ j 2 \big [\CT j 1 \big ]
  + \big[\CZ j 1\big]^3 \big [\CT j 1 \big ] \,.
  \label{eq:DeltaE4}
\end{multline}
I calculate $\Delta E_{j,\thr}^{(4)}$ only up to order $1/L^6$ in the
volume expansion. From Ref.~\cite{Hansen:2015zta}
the leading $1/L$ behavior of the terms in Eq.~(\ref{eq:DeltaE4}) is
known to be
\begin{equation}
\begin{gathered}
\CZ j 2 \sim 1/L^2\,, \qquad
\CZ j 1 \sim \big[\CT j 1\big] \sim 1/L^3\,,\\
\big[\CT j 2 \big] \sim 1/L^4\,, \qquad
\big[\CT j 3\big] \sim 1/L^5\,.
\label{eq:CZj2}
\end{gathered}
\end{equation}
Explicit examples are given below.
Thus the only contributions that must be kept are
\begin{equation}
\Delta E_{j,\thr}^{(4)} = - \CT j 4 
+ \CZ j 2 \big [\CT j 2 \big ]
 + \CZ j 3 \big [\CT j 1 \big ] 
+ {\cal O}(1/L^7)\,.
  \label{eq:DeltaE4fin}
\end{equation}
Since $\CZ j 2$, $\big[\CT j 2 \big]$ and $\big[\CT j 1\big]$ are
determined in Ref.~\cite{Hansen:2015zta}, the only new quantities
needed here are the $1/L^6$ contributions to $\big[\CT j 4\big]$ 
and the $1/L^3$ contributions to $\CZ j 3$. For both quantities these 
are the leading contributions in the $1/L$ expansion.

\bigskip
I now describe the results that  I aim to check.
The threshold expansion for the energy shift for two particles 
follows from the general formalism of Refs.~\cite{Luscher:1986n2,Luscher:1991n1}.
It is worked out through $\cO(1/L^5)$ in Ref.~\cite{Luscher:1986n2}
and the $1/L^6$ term is given in Ref.~\cite{Hansen:2015zta}.
The result is
\begin{equation}
\Delta E_{2,\thr} = \frac{4 \pi a}{m L^3} \left\{
1 - \left(\frac{a}{\pi L}\right)\cI 
+ \left(\frac{a}{\pi L}\right)^2(\cI^2-\cJ)
+ \left(\frac{a}{\pi L}\right)^3\left[-\cI^3 + 3\cI \cJ - \cK\right]
+ \frac{2\pi r a^2}{L^3}
- \frac{\pi a}{m^2 L^3}  \right\} + \cO(L^{-7})\,,
\label{eq:DE2gen}
\end{equation}
with $a$ the scattering length (defined to be positive for repulsive interactions), 
$r$ the effective range,
and $\cI$, $\cJ$, $\cK$ are known sums over functions of integer vectors
(see Appendix~\ref{app:sumint}).
The result for the three-particle threshold energy is~\cite{Hansen:2016fzj}\footnote{%
In the initial published version the coefficient of $\cK$ was $-9$, but this was corrected
in an Erratum to $+15$~\cite{Hansen:2016fzj}.}
\begin{multline}
\Delta E_{3,\thr} =
\frac{12 \pi a}{m L^3} \Bigg\{
1 - \left(\frac{a}{\pi L}\right)\cI + \left(\frac{a}{\pi L}\right)^2(\cI^2+\cJ)
+ \frac{64 \pi^2 a^2 \cC_3}{mL^3}
+ \frac{3\pi a}{m^2 L^3}
+ \frac{6\pi r a^2}{L^3} 
\\
+ \left(\frac{a}{\pi L}\right)^3\left[
-\cI^3 +\cI\cJ + 15 \cK % -9\cK 
+ c_L \log(N_{\rm cut}) + \cC_F + \cC_4 + \cC_5
\right]
\Bigg\}
- \frac{\Mthr}{48 m^3 L^6} + \cO(L^{-7})
\,,
\label{eq:DE3gen}
\end{multline}
where $N_{\rm cut}=mL/(2\pi)$,
$c_L= 16 \pi^3 (\sqrt3-4\pi/3)$,
and $\cC_F$, $\cC_3$, $\cC_4$ and $\cC_5$ are sums over
integer vectors that are defined and evaluated in Ref.~\cite{Hansen:2016fzj}.
The new amplitude entering at $\cO(1/L^6)$ is the divergence-free three-to-three
threshold amplitude $\Mthr$, which begins at $\cO(\lambda^2)$ in perturbation theory.
The numerical values of $\cC_3$, $\cC_4$ and $\cC_5$ depend on the choice
of UV cutoff, but this dependence cancels with that of $\Mthr$.
This cancelation is necessary because $\Delta E_{3,\thr}$ is a physical quantity.

Since $a$ and $\lambda$ are proportional [Eq.~(\ref{eq:lambdadef})], the dependence
of $\Delta E_{j,\thr}$ on $\lambda$
is manifest except for the terms involving $r$ and $\Mthr$.
To make the perturbative expansion clearer I rewrite $r$, using its definition, as
\begin{equation}
32 \pi m^3 r a^2 \equiv - \lambda - 2 \cK'_{2,s,\thr}
\,,
\end{equation}
where
\begin{equation}
\cK'_{2,s,\thr} \equiv m^2 \frac{d \cK_{2,s}}{d q^2} \bigg|_{\thr}
\,,
\label{eq:Kprime}
\end{equation}
Here $\cK_{2,s}$ is the two-particle s-wave K matrix, and $q$ is the momentum of each
particle in the two-particle CM frame.
The perturbative series for $\cK'_{2,s,\thr}$ and $\Mthr$ both
begin at $\cO(\lambda^2)$:
\begin{equation}
\cK'_{2,s,\thr} = \sum_{n=2}^\infty \lambda^n \cK'^{(n)}_{2,s,\thr}\,,
\qquad
\Mthr = \sum_{n=2}^\infty \lambda^n \Mthr^{(n)}
\,.
\end{equation}
Combining these results, the predictions above imply that the
fourth-order terms are 
\begin{align}
\Delta E_{2,\thr} ^{(4)} &= \frac{1}{2^{18}\pi^6 m^5 L^6} 
\left[ -\cI^3 + 3\cI \cJ - \cK \right]
- \frac{\cK'^{(3)}_{2,s,\thr}}{2^6 m^5 L^6}
+ \cO(L^{-7}) \,,
\label{eq:DE24}
\\
\Delta E_{3,\thr}^{(4)} &= \frac{3}{2^{18}\pi^6 m^5 L^6} 
\left[ -\cI^3 +\cI\cJ + 15 \cK % -9\cK 
+ c_L \log(N_{\rm cut}) + \cC_F + \cC_4 + \cC_5
\right]
- \frac{9 \cK'^{(3)}_{2,s,\thr}}{2^6 m^5 L^6}
- \frac{\Mthr^{(4)}}{48 m^3 L^6} + \cO(L^{-7})
\,.
\label{eq:DE34}
\end{align}
In order to separate out effects that are particular to the three-particle case, 
it is convenient to consider the difference
\begin{equation}
\Delta_{32} = \Delta E_{3,\thr} - 9 \Delta E_{2,\thr} 
= \sum_{n=2}^\infty \lambda^n \Delta_{32}^{(n)} 
\,,
\end{equation}
for which the fourth-order coefficient is predicted to be
\begin{equation}
\Delta_{32}^{(4)} =
\frac{3}{2^{18}\pi^6 m^5 L^6}
\left[
2\cI^3 - 8\cI\cJ+18\cK + c_L \log(N_{\rm cut}) + \cC_F + \cC_4 + \cC_5
\right]
- \frac{\Mthr^{(4)}}{48 m^3 L^6} + \cO(L^{-7})
\,.
\label{eq:DE324}
\end{equation}
Note that the effective range has canceled from this expression.

%In the following I focus on the $L^{-6}$ contributions
%to Eqs.~(\ref{eq:DE24}) and (\ref{eq:DE324}) and 
%in general do not show explicitly the $\cO(L^{-7})$ corrections.

%The main purpose of the present note is to check the explicit $a^4$ terms in
%the three-particle energy shift, as well as the $\lambda^4$ contributions to
%$\Mthr$, since the latter involve a new type of subtraction of divergences not
%present at lower orders.

To motivate the definition of $\Delta_{32}$, I recall from
Ref.~\cite{Hansen:2015zta} that the three-particle correlators
can be split into a ``connected'' part,
containing contributions in which the Feynman diagram connects all three particles, 
a ``disconnected'' part in which one particle is a spectator (possibly having
self-energy insertions) and the other two are connected, 
and the fully disconnected remainder (which does not lead to power-law finite-volume effects).
Since there are three possible two-particle pairs in a three-particle system,
the following relations hold for all $n$,
\begin{equation}
C^{(n)}_{3,\thr,\disc}(0) = 3\, \CZ 2 n\,, 
\qquad
\partial_\tau C^{(n)}_{3,\thr,\disc}(0) = 3\, \CT 2 n\,.
\end{equation}
As noted in Ref.~\cite{Hansen:2015zta}, for $n=1$ and $2$,
connected contributions to $C_3$ do not begin until $\cO(1/L^6)$,
so that
\begin{align}
\CZ 3 n &= 
%C^{(n)}_{3,\thr,\disc}(0) + \cO(1/L^6) = 
3\, \CZ 2 n + \cO(L^{-6}) & (n=1,2) \,.
\label{eq:CZ32rel}
\end{align}
while the low order contributions to the connected part of $\CZ 3 {n}$ satisfy
\begin{equation}
\partial_\tau C^{(1)}_{3,\thr,\conn}(0) = 0 \,,\qquad
\partial_\tau C^{(2)}_{3,\thr,\conn}(0) = \cO(L^{-6})\,.
\end{equation}
Combining these results yields
\begin{equation}
\Delta_{32}^\four = -\partial_\tau C^{(4)}_{3,\thr,\conn}(0) 
+ 3 \, \CZcon 3 3 \CT 2 1+ 6\, \CT 2 4  
+ \cO(L^{-7})
\,.
\label{eq:D32calc}
\end{equation}
showing that several two-particle quantities have canceled in the difference.

To summarize the previous discussion,
the new quantities that are needed to determine
$\Delta E_{3,\thr}^{(4)}$ are
$\CZ 2 3$, $\CT 2 4$, $\CTcon 3 3$ and $\CTcon 3 4$.
Once these quantities have been calculated it requires no extra work to 
determine the result for $\Delta E_{2,\thr}^\four$.
Having done so, it is convenient to 
consider $\Delta_{32}^\four$ instead of $\Delta E_{3,\thr}^{(4)}$.
Breaking up the calculation in this way also proved useful in practice for
tracking down errors.

\bigskip
The calculation of the finite-volume correlation functions proceeds as in
Ref.~\HSPT. Propagators are written in their time-momentum form,
i.e. $\exp(-|\Delta t| \omega_p)/(2 \omega_p)$ with $\omega_p=\sqrt{m^2 + p^2}$
and $p=|\vec p|$. The integrals over the vertex times, $\tau_i$, are then 
straightforward but tedious.\footnote{%
I use Mathematica to do these integrals, and have found that doing more
than two integrals at once can lead to incorrect results.
%due to issues with
%keeping track of absolute values $|\tau_i-\tau_j|$ in the exponents.
Thus all integrals are done stepwise, with numerical checks at each stage.}
This leaves a sum over momenta of a summand that is, in general, 
quite complicated. For the sake of brevity, I do not display these summands
except in a few cases.\footnote{%
Expressions for all integrands or summands are available upon request from the author.}
The sums are always UV finite after inclusion of counterterms.
There are up to three loop-momenta in the diagrams considered.

At this stage the sums are replaced by integrals 
plus a volume-dependent difference.
The general analysis of Refs.~\cite{Luscher:1986n2,Kim:2005,Hansen:2015zta},
implies that the sum-integral difference is exponentially suppressed in $L$
(typically as $e^{-m L}$) except for loops in which intermediate particles can go on shell.
Such loops have summands that diverge in the IR, and
the results collected in Appendix~\ref{app:sumint} can be used to pull out the
dominant volume dependence.
What is left is a finite integral that is, in the present calculation,
at most of two-loop order. Such integrals can easily be evaluated numerically.
The tests presented here also requires a two-loop calculation of the scattering
length and the three-particle subtracted threshold amplitude, $\Mthr$.
These are infinite-volume quantities where the calculations are most easily
done using standard momentum-space Feynman rules and dimensional regularization.
The calculations are outlined in Appendix~\ref{app:K}.

\section{Determining $\Delta E_{2,\thr}^{(4)}$.}
\label{sec:DE24}

In this section I calculate the $\lambda^4$
contribution to $\Delta E_{2,\thr}$.
Given the form of the expected answer, I write
\begin{equation}
\Delta E_{2,\thr}^\four = \frac{a_{2}^\four}{2^{18} \pi^6 m^5 L^6} + \cO(L^{-7})\,,
\label{eq:a24def}
\end{equation}
and quote results for $a_2^\four$.

I begin by collecting results from Ref.~\cite{Hansen:2015zta} that
are needed in order to evaluate 
$\Delta E_{2,\thr}^{(4)}$ using Eq.~(\ref{eq:DeltaE4fin}):\footnote{%
The results in Eq.~(\ref{eq:CZT22}) look different from those given
in Eqs.~(27)  and (28) of  Ref.~\cite{Hansen:2015zta} 
because here I expand in the renormalized rather than the bare coupling.
In particular, terms proportional to $A_2/L^3$ 
present in Ref.~\cite{Hansen:2015zta} are canceled here by 
contributions from the $\mathcal O(\lambda^2)$ counterterm.}
\begin{align}
\CZ 2 1 &= -\frac1{16 m^3 L^3}\,,
&\CT 2 1 = - \frac{1}{8 m^2 L^3} \,, 
\label{eq:CZT21}
\\
\CZ 2 2 &= \frac{\mathcal I}{2^8 \pi^2 m^3 L^4} + {\cal O}(L^{-6})\,,
&\CT 2 2 = - \frac{\mathcal J}{2^{10} \pi^4 m^2 L^2}
+ {\cal O}(L^{-3})\,,
%+ \frac{1}{2^9 \pi m^3 L^3} + {\cal O}(1/L^4)\,,
\label{eq:CZT22} 
\end{align}
Using these results one can immediately determine the 
$\CZ 2 2 \CT 2 2$ contribution in Eq.~(\ref{eq:DeltaE4fin}), leading to\footnote{%
Here and in the following I use the proper superset symbol $\supset$ to indicate
individual contributions to quantities (with the quantity here being $a_2^\four$).
In the course of the calculation I determine all contributions of the desired order
and collect them in a final result.}
\begin{equation}
a_2^\four \supset - I J
\label{eq:DE24term22}
\end{equation}
What remains is to calculate $\CZ 2 3$ and $\CT 2 4$.

\subsection{Calculating $\CZ 2 3$}
\label{sec:CZ23}

\begin{figure}[tbh]
\begin{center}
\vskip -.2truein
\includegraphics[scale=0.5]{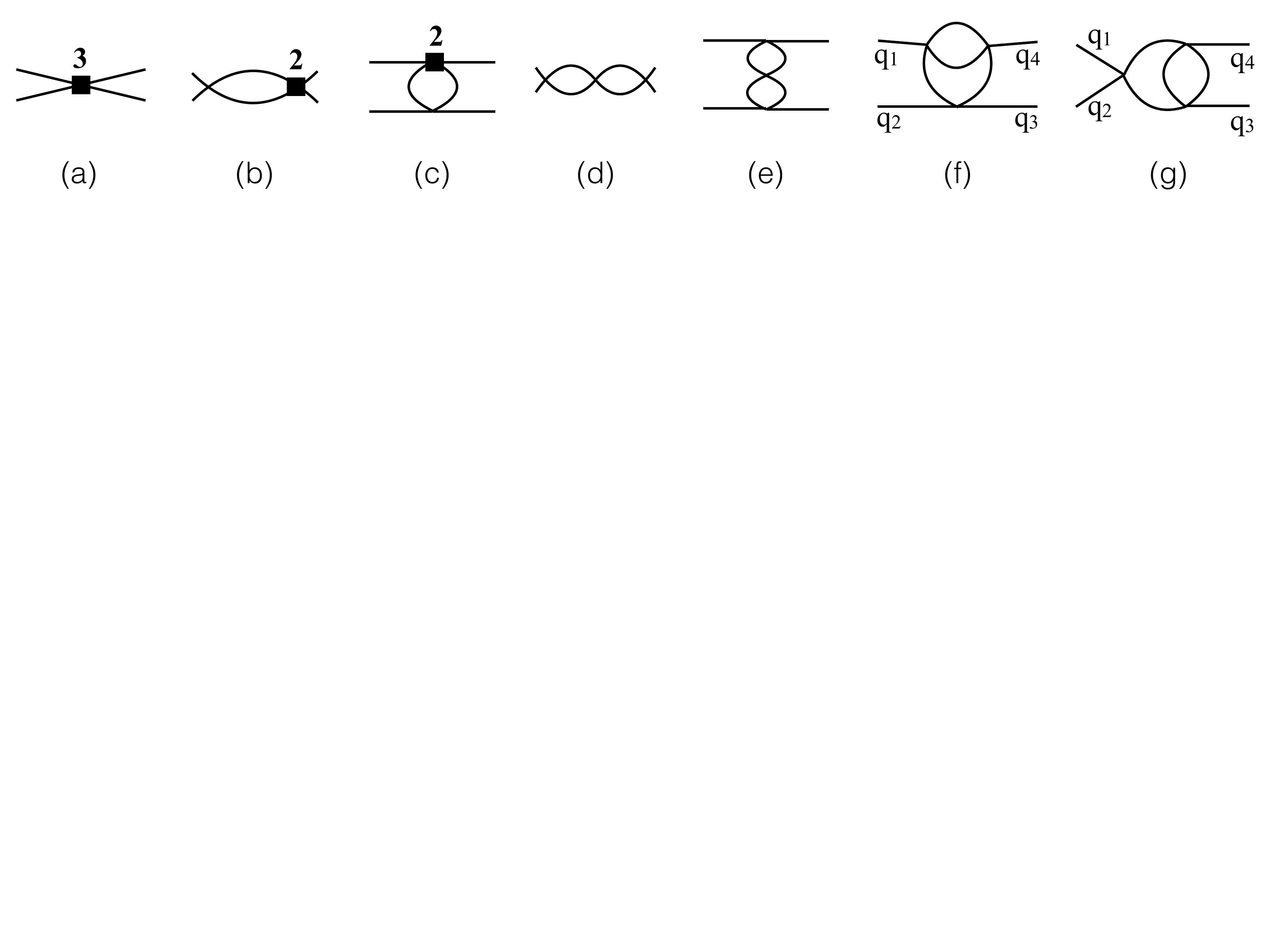}
\vskip -4.2truein
\caption{Feynman diagrams contributing to $C_2(\tau)$ at $\mathcal O (\lambda^3)$. 
Solid squares are vertex counterterms, with the number indicating the power of $\lambda$.
External particles have zero three-momentum. 
Diagrams related by vertical or horizontal reflection are not shown explicitly.
Interpreted as contributions to the infinite-volume scattering amplitude, 
these are the diagrams required to calculate the two-loop counterterms 
in Appendix~\ref{app:counter}, and the two-loop K matrix in Appendix~\ref{app:K}.
The momentum labels in (f) and (g) are used in the latter calculation.
Time runs from left to right in both applications.
}
\label{fig:C2lam3}
\end{center}
\end{figure}

The diagrams needed to calculate $\CZ 2 3$ are shown in Fig.~\ref{fig:C2lam3}.
%Except for the addition of counterterms, these are the same diagrams
%as those used in Ref.~\cite{Hansen:2015zta} to calculate $\big[\CT 2 3\big]$.
Since $\CZ 2 3$ appears in Eq.~(\ref{eq:DeltaE4fin}) multiplied by
$\CT 2 1={\cal O}(L^{-3})$,
$\CZ 2 3$ itself is needed only up to ${\cal O}(L^{-3})$.

\subsubsection{SS diagram}

I begin by determining the contribution of Fig.~\ref{fig:C2lam3}(d), together with 
the $A_{2s}$ contribution to Fig.~\ref{fig:C2lam3}(b), 
plus its horizontal reflection, and the $A_{3ss}$ contribution to Fig.~\ref{fig:C2lam3}(a).
I label the left- and right-hand loop momenta $p$ and $q$, respectively.
If both momenta vanish then the contribution is of $\cO(L^{-9})$, 
well below the order of interest.
Contributions of $\cO(L^{-3})$ do arise, however, if one or both momenta are nonzero.

Consider first the case in which one momentum vanishes, say $q$.
Then it is possible for all three time integrals to give factors of $1/p^2$,
each arising from integrals of the form
\begin{equation}
\int_{\tau_i}^{\tau_k} d\tau_j \, e^{-(\tau_k-\tau_j) 2(\omega_p-m)}
\propto \frac{1-e^{-(\tau_k-\tau_i) 2 (\omega_p-m)}}{2(\omega_p-m)} \sim \frac{1}{p^2}\,.
\end{equation}
%This can occur because two-particle cuts can be made through both loops.
Explicit evaluation 
(including a factor of $2$ from the fact that either loop momentum can vanish)
yields
\begin{align}
\CZ 2 3 &\supset  - \frac{1}{2^8m^3 L^9} \sum_{\vec p\ne 0} \frac1{p^6} 
\left[1 + \mathcal O(p^2) \right]
\\
&= -\frac{\mathcal K}{2^{14} \pi^6 m^3 L^3}\left[1 + \mathcal O(L^{-1}) \right]
\,,
\end{align}
Here I have kept only the most singular part of the summand, since less singular terms
contribute at subleading order in $L^{-1}$. To obtain the second line
I have used Eq.~(\ref{eq:Kdef}).
Note that, although a sum over $p$ usually absorbs a single factor of $L^{-3}$
(in order to become an integral), the presence of the $1/p^6$ IR divergence means
that a factor of $L^{-6}$ is absorbed. This brings the contribution up to the
desired order.

If both loop momenta are nonvanishing, the summand is simple and so I display the
complete result:
\begin{align}
\CZ 2 3 &\supset \frac{1}{2^9 m^3 L^3} \Bigg\{
\frac1{L^3}\sum_{\vec p\ne 0}\frac{m^2}{\omega_p p^4}
\left[\frac1{L^3}\sum_{\vec q\ne 0} \!-\!\int_q\right]\frac1{\omega_q q^2}
+ (p\leftrightarrow q)
- \frac12 \left[\frac1{L^3}\sum_{\vec p\ne 0}\! -\!\int_p\right]\frac1{\omega_p p^2}
 \left[\frac1{L^3}\sum_{\vec q\ne 0} \!-\!\int_q\right]\frac1{\omega_q q^2}\Bigg\}
\label{eq:C30SS}
\\
&=
\frac{\cI\cJ}{2^{14} \pi^6 m^3 L^3}\left[1 + \mathcal O(L^{-1}) \right]
\,.
\end{align}
To obtain the second line  I have  used Eqs.~(\ref{eq:Idef}) and (\ref{eq:Jdef}).
Note that the maximal degree of IR divergence is the same as for when
one momentum vanishes, but now the divergence is split between $p$ and $q$. 
The final term in Eq.~(\ref{eq:C30SS}) has a lower degree of IR divergence,
and gives a subleading contribution.

\subsubsection{Remaining diagrams}

The TT diagram, Fig.~\ref{fig:C2lam3}(e), combines with  
the $A_{2t}$ contribution to Fig.~\ref{fig:C2lam3}(c),
and the $A_{3tt}$ contribution to Fig.~\ref{fig:C2lam3}(a).
In this case, the absence of physical cuts allows the replacement of sums
with integrals. The combined integrand, including
counterterms, is UV and IR convergent:
\begin{align}
\CZ 2 3 &\supset - \frac1{2^{12} \pi^4 m^3 L^3} I^{\rm TT}\,,
\label{eq:TT}
\\
I^{\rm TT} &=8 \pi^4 \int_p \int_q 
\frac{m^3}{\omega_p^3 (\omega_p+m) \omega_q^3(\omega_q+m)(\omega_p+\omega_q)}
= \frac{2(\pi-3)}{3}
\,.
\label{eq:ITT}
\end{align}

The SU diagram of Fig.~\ref{fig:C2lam3}(f) combines with 
the $A_{2s}+A_{2u}$ contribution to Fig.~\ref{fig:C2lam3}(c),
and the $A_{3su}$ contribution to Fig.~\ref{fig:C2lam3}(a).
Again, sums can be replaced by integrals, leading to
\begin{align}
\CZ 2 3 &\supset - \frac1{2^9 \pi^4 m^3 L^3} I^{\rm SU}\,,
\qquad I^{\rm SU} = {0.0396563}
\,.
\label{eq:ISU}
\end{align}
I only give the result of numerical integration,
since the integrand is long and uninformative.

Finally, the ST diagram, Fig.~\ref{fig:C2lam3}(e), combines
the $A_{2t}+A_{2u}$ contribution to Fig.~\ref{fig:C2lam3}(b),
and the $A_{3st}$ contribution to Fig.~\ref{fig:C2lam3}(a),
together with their horizontal reflections.
The total contribution only scales as $1/p^2$ in the IR, with no IR divergence in $q$.
Thus both sums can be replaced by integrals up to corrections of relative size $L^{-1}$,
leading to
\begin{align}
\CZ 2 3 &\supset - \frac{I^{\rm ST}}{2^9 \pi^4 m^3 L^3} 
\,,\qquad
I^{\rm ST} = 0.099447 % \frac{0.198893}{2} 
\,.
\label{eq:ST}
\end{align}
%where again I do not show the explicit form of the integrand.

\subsubsection{Total contribution to  $\Delta E_{2,\thr}^\four$}

\begin{comment}
The total contribution is thus
\begin{align}
\CZ 2 3 &\supset
\frac{(\mathcal I\mathcal J - \mathcal K)}{2^{14} \pi^6 m^3 L^3} 
-\frac{0.3018}{2^{10} \pi^4 m^3 L^3}
\,.
\end{align}
\end{comment}
Multiplying the above results by $\CT 2 1$ from Eq.~(\ref{eq:CZT21})  yields
\begin{align}
a_2^\four&\supset -2\cI \cJ + 2 \cK + 2^6 \pi^2 (I^{\rm ST}+\tfrac18 I^{\rm TT}  +I^{\rm SU})
\,.
\label{eq:DECT3}
\end{align}

\subsection{Contribution of $\CT 2 4$}
\label{sec:CT24}

\begin{figure*}[tbh]
\begin{center}
\includegraphics[scale=0.5]{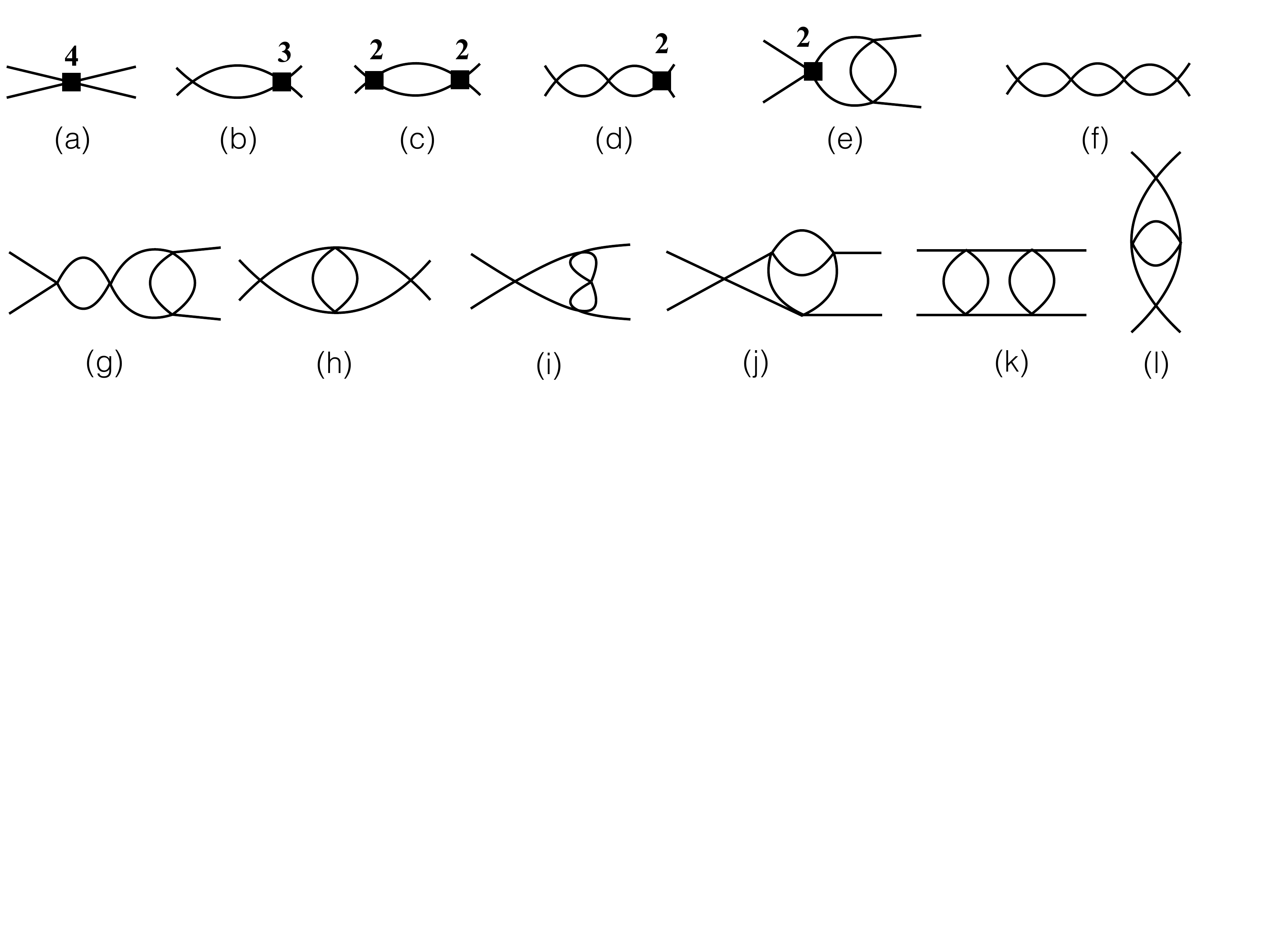}
\vskip -3 truein
\caption{Subset of Feynman diagrams contributing $C_2(\tau)$
at $\mathcal O (\lambda^4)$. Notation as in Fig.~\ref{fig:C2lam3}.
Figures (b)-(k) show all the diagrams (aside from reflections, and additional placements of
counterterms)
for which there is a two-particle cut.
Figure (l) shows a single example of the many diagrams without such cuts.
See text for further discussion.}
\label{fig:C2lam4}
\end{center}
\end{figure*}

In this section I calculate the contribution to $a_2^\four$ from $\CT 2 4$.
%using $\Delta E_{2,\thr}^{(4)} \supset - \CT 2 4$.
%\begin{equation}
%a_2^\four \supset - \left[\CT 2 4\right] 2^{18} \pi^6 m^5 L^6
%\,.
%\end{equation}
A  large number of diagrams contribute to $\CT 2 4$, a subset
of which is shown in Fig.~\ref{fig:C2lam4}.
I first describe some general properties of the contributions of these diagrams
{\em if all loop momenta are nonzero}.
In this case, a term linear in $\tau$
arises only from a configuration in which all vertices lie close in time
and are integrated as a group over the full time interval. Configurations
in which some vertices are separated by $\cO(\tau)$ are
exponentially suppressed.
Thus the contribution to $\CT 2 4$ arises from two particles at rest propagating
freely between $0$ and $\tau$, except for a single quasilocal interaction.
From this one can show that, as $L\to\infty$,  the leading volume dependence
of $\CT 2 4$ has
the form $c/(8 m^2 L^3) $, where $c$ is the contribution
of the diagram (now viewed as an infinite-volume scattering diagram)
to the scattering amplitude at threshold, $\cM_{2,\thr}$.\footnote{%
Indeed, this is exactly the form that arises at tree level, where
$\lambda \CT 2 1= -\lambda/(8 m^2 L^3)$.}
Note that, when taking the $L\to\infty$ limit, all sums are replaced by integrals,
$(1/L^3)\sum_{\vec p} \to \int_p$.

\begin{comment}
To show this in detail one uses the results of Ref.~\HSQCb\ to relate
the correlation function to a ``finite-volume  scattering amplitude" whose
infinite-volume limit is the physical scattering amplitude.
This relation requires amputation, which is what is achieved by the overall
factors in the definition of $C_2(\tau)$ and the $-1/(8 L^3)$ factor in $\CT 2 4$.
\end{comment}

This result has two important consequences. The first is practical: it
allows the determination of the integrand of $\cM_{2,\thr}$
from the summand appearing in $\CT 2 4$ on a diagram by diagram basis.
The prescription is simply to multiply the summand by $8 m^2 L^{12}$.
Here the factor of $8 m^2 L^3$ noted above is multiplied by
$L^9$ due to the conversion of three momentum sums into integrals.
I use this result to calculate the counterterms quoted in Appendix~\ref{app:counter}.

The second consequence is that 
the constant $c$ vanishes when each three-loop diagram is combined with
the corresponding counterterms. This is because
the $\cO(\lambda^4)$ contributions to $\cM_{2,\thr}$ vanish
in the renormalization scheme I use. (Indeed, the only contribution is
of $\cO(\lambda)$.)
Since $c$ is obtained by replacing momentum sums with integrals, 
it follows that all finite-volume corrections arise from sum-integral differences.
I stress again that this argument holds for the case in which
all loop momenta are nonvanishing.

From this result follows a key simplification in the calculation of $\CT 2 4$:
only diagrams containing two-particle cuts can contribute.
These are the diagrams shown in Figs.~\ref{fig:C2lam4}(f)-(k).
For diagrams without such cuts, such as Fig.~\ref{fig:C2lam4}(l),
sum-integral differences are exponentially suppressed
and do not lead to power law volume dependence.
Furthermore, for diagrams without cuts,
the cases in which loop momenta vanish do not require separate consideration,
as there are no IR divergences.

For the diagrams with two-particle cuts, one must also consider the cases in
which one or more loop momenta vanish. In these cases the summands
are not related to integrands of $\cM_{2,\thr}$, do not vanish, and
must be calculated explicitly. If one loop momentum vanishes, then the
contribution is of $\cO(L^{-6})$ if the other loop sums are replaced by integrals.\footnote{%
It is possible in principle that IR divergences could reduce the power of $1/L$, but this
does not occur in practice.}
If two loop momenta vanish then the contribution begins at $\cO(L^{-9})$ and
can be raised to the desired $L^{-6}$ behavior only if there is a $1/p^6$ IR
divergence. This only occurs for Fig.~\ref{fig:C2lam4}(f).
If all three loop momenta vanish, then the contribution is of $\cO(L^{-12})$ and
can be dropped.

I now consider Figs.~\ref{fig:C2lam4}(f)-(k) in turn,
calling them, respectively,
the SSS, SST, STS, STT, SSU and TST diagrams.

\subsubsection{SSS diagram}

Figure.~\ref{fig:C2lam4}(f) combines with the $A_{4sss}$ part of
Fig.~\ref{fig:C2lam4}(a), the $A_{3ss}$ part from Fig.~\ref{fig:C2lam4}(b),
the $A_{2s}^2$ part from Fig.~\ref{fig:C2lam4}(c) and the $A_{2s}$ part
from Fig.~\ref{fig:C2lam4}(d). 
I find
\begin{align}
\CT 2 4 &\supset \frac{\cI^3 - 6\cI \cJ + 3\cK}{2^{18}\pi^6 m^5 L^6}
\,,
\label{eq:SSS}
\end{align}
with the three terms arising, respectively, from having zero, one and two
nonzero loop momenta.
%\footnote{%
%
%If all three loop momenta in Fig.~\ref{fig:C2lam4}(f) vanish, the contribution to
%$\CT 2 4$ is proportional to $1/L^{12}$, well beyond the order of interest.}
%

\subsubsection{SST diagram}

Next I consider Fig.~\ref{fig:C2lam4}(g), together with 
the $A_{4sst}$ contribution to Fig.~\ref{fig:C2lam4}(a), the
$A_{3st}$ contribution to Fig.~\ref{fig:C2lam4}(b), the $A_{2s} (A_{2t}+A_{2u})$ contribution
to Fig.~\ref{fig:C2lam4}(c), the $A_{2t}+A_{2u}$ contribution to Fig.~\ref{fig:C2lam4}(d), 
and the $A_{2s}$ contribution to Fig.~\ref{fig:C2lam4}(e).

The sum over the momenta in the rightmost loop can always be converted to
an integral since the summand is nonsingular.
%If both the other loops have vanishing momenta the contribution is of $\mathcal O(1/L^9)$
%and can be dropped.
Thus at most one of the remaining loops can have vanishing momenta. 
I describe the calculations in some detail.

If all three loop momenta are nonvanishing, contributions arise from
(a) a sum-integral difference on the left loop (with the other loops integrated),
(b) a sum-integral difference on the central loop (with other loops integrated), and
(c) sum-integral differences on left and central loops (with the rightmost loop integrated).
I find by explicit calculation that the summands/integrands for the first two cases vanish
identically.
The explicit expression for case (c) is (including the horizontal reflection):
\begin{align}
\CT 2 4 &\supset \frac1{2^{10} L^3} 
\left\{\left[\frac1{L^3} \sum_{\vec p \ne 0}-\int_p\right]
 \frac{1}{\omega_p\vec p^2}\right\}
\left\{ \left[\frac1{L^3}\sum_{\vec k\ne 0} -\int_k\right]
\frac{1}{\omega_k\vec k^2} \int_q \frac{f^{\rm SST}(\vec k,\vec q)}{\omega_q^3} 
\right\}\,,
\label{eq;C24g2}
\\
f^{\rm SST}(\vec k, \vec q)& =
-1 +
\frac{2\omega_q^2 (\omega_k+W_{qk})}{\omega_{qk}(W_{qk}^2-1)}
\,,
\end{align}
where $W_{qk}=\omega_q+\omega_k+\omega_{qk}$ and 
$\omega_{qk}^2 = m^2 +(\vec q + \vec k)^2$.
A key result is that $f^{SST}(0, \vec q)=0$.
Using Eq.~(\ref{eq:Idef}), one sees that the expression in the left-hand curly braces is
proportional to $1/L$, while that in the right-hand curly braces is proportional to $1/L^3$,
so that the overall contribution to $\CT 2 4$ is proportional to $1/L^7$ and can be dropped.

If the leftmost loop momentum vanishes, it turns out that the central loop has an integrable
$1/k^2$ IR divergence. Replacing the central momentum sum with an integral
[valid up to corrections of $\cO(L^{-1})$], and including the horizontal reflection, yields the
result
\begin{align}
\CT 2 4 &\supset \frac{I^{\rm SST0}}{2^{13} \pi^4 m^5 L^6}\,, \quad
I^{\rm SST0}=0.19889\,.
\label{eq:SST0}
\end{align}
Here $I^{\rm SST0}$ is a UV convergent two-loop integral of a lengthy expression
that I evaluate numerically.
I note that the relation $I^{\rm SST0}=2 I^{\rm ST}$ holds numerically.

If the central loop momentum vanishes, then,
if the left-hand momentum sum is replaced by an integral, the result vanishes
identically. It follows that there is no $\cO(L^{-6})$ contribution to $\CT 2 4$.

\subsubsection{STS diagram}

Figure~\ref{fig:C2lam4}(h) combines with
Fig.~\ref{fig:C2lam4}(d) 
(in which the $A_{2t}+A_{2u}$ counterterm is placed on the middle vertex), 
as well as the $A_{3st}$ contribution to Fig.~\ref{fig:C2lam4}(b)
(together with its reflection)
and the $A_{4sts}$ contribution to Fig.~\ref{fig:C2lam4}(a).
%If both outer-loop momenta vanish, then the contribution to $\CT 2 4$ 
%is proportional to $1/L^9$ and can be dropped.

The central loop sum can always be converted to an integral without power-law volume
corrections.
If both of the outer loop momenta are nonzero, I find
(with $\vec p$ and $\vec k$ the momenta in the outer loops, and $\vec q$ the
central momentum)
\begin{align}
\CT 2 4 &\supset 
\frac1{2^{11} m^2 L^3}  
\left[\frac1{L^3} \sum_{\vec p \ne 0}- \int_p\right]
\left[\frac1{L^3} \sum_{\vec k\ne 0}-\int_k\right]  \frac{1}{\vec p^2 \vec k^2}
\int_q f^{\rm STS1}(\vec p, \vec k, \vec q) 
\nonumber \\
&\quad+
\frac1{2^{10} m^2 L^3}  
\left[\frac1{L^3} \sum_{\vec p \ne 0}- \int_p\right] \frac{1}{\vec p^2 }
\int_k \int_q f^{\rm STS2}(\vec p, \vec k,\vec q)
% result for integrand without counterterms
%f^{(h1)}(\vec p, \vec k, \vec q) &=
%\frac{2(W_{pqk}^2 +W_{pqk}(\omega_p+\omega_k) + 2 \omega_p\omega_k -2)}
%{\omega_p \omega_k \omega_{pq}\omega_{kq} W_{pqk}(W_{pqk}^2-4)}
%where $W_{pqk}=\omega_p+\omega_k + \omega_{pq}+\omega_{kq}$.
\,.
\label{eq:C24h4}
\end{align}
The relevant properties of the functions $f^{\rm STS1}$ and $f^{\rm STS2}$ will be given below.
To study the first term in Eq.~(\ref{eq:C24h4}), I introduce
\begin{equation}
g^{\rm STS1}(\vec p^{\,2}, \vec k^{\, 2}, \vec p\cdot \vec k) 
= \int_q f^{\rm STS1}(\vec p, \vec k, \vec q)
\,,
\end{equation}
where the form of the arguments of  $g^{ \rm STS1}$ is determined by rotation invariance.
Generalizing the analysis leading to Eq.~(\ref{eq:Idef}) gives
\begin{multline}
\left[\frac1{L^3} \sum_{\vec p \ne 0}- \int_p\right]
\left[\frac1{L^3} \sum_{\vec k\ne 0}-\int_k\right] 
 \frac{g(\vec p^{\,2},\vec k^{\,2}, \vec p \cdot \vec k)}{\vec p^2 \vec k^2}
 = \left(\frac{\mathcal I}{4\pi^2 L}\right)^2 g(0,0,0)
 -  \frac{\mathcal I}{4 \pi^2 L^4}
 \left[\frac{\partial}{\partial p^2}+\frac{\partial}{\partial k^2}\right] g\Bigg|_{\vec p=\vec k=0}
 + {\cal O}(1/L^6)
\,.
\end{multline}
%Here I assume that the sums and integrals are regulated in the UV.
Using the result
$g^{\rm STS1}(0,0,0) = 0$,
which follows from the renormalization condition,
I find that the leading finite-volume term is proportional to $L^{-7}$.

Turning to the second term in Eq.~(\ref{eq:C24h4}), I introduce
\begin{equation}
g^{\rm STS2}(\vec p^{\,2}) = \int_k\int_q f^{\rm STS2}(\vec p, \vec k,\vec q)
\,,
\end{equation}
which is a function of $\vec p^{\,2}$ by rotation invariance.
Using
\begin{equation}
g^{\rm STS2}(0)= 0\,, \qquad
{g^{\rm STS2}}'(0) = \frac{I^{\rm STS}}{8 m^3 \pi^4}\,, \quad
I^{\rm STS} = -0.37115\,,
\end{equation}
together with Eq.~(\ref{eq:Idef}), the second term in
Eq.~(\ref{eq:C24h4}) contributes
\begin{align}
\CT 2 4 &\supset 
%- \frac1{2^{10} m^2 L^3}  \frac{ g'^{(STS2)}(0) }{ L^3}
\frac{-I^{\rm STS}}{2^{13} \pi^4 m^5 L^6} 
\,.
\label{eq:STS}
\end{align}
%The vanishing of $g^{(STS2)}(0)$ implies that there is no $1/L^4$ term, as expected.

If one or other of the outer momenta vanishes, then I find
\begin{align}
\CT 2 4 &\supset \frac{I^{\rm STS0}}{2^{13} \pi^4 m^5 L^6}\,,\qquad
I^{\rm STS0} =I^{\rm SST0}
\,,
\label{eq:STS0}
\end{align}
where the latter equality holds to numerical precision.

\subsubsection{STT diagram}

Figure~\ref{fig:C2lam4}(i) combines with
the $A_{2t}$ contribution to Fig.~\ref{fig:C2lam4}(e) (with the counterterm
on the two right-hand vertices), the $A_{3tt}$ contribution to
Fig.~\ref{fig:C2lam4}(b) and the $A_{4stt}$ contribution to Fig.~\ref{fig:C2lam4}(a).

The sums over momenta in the right-hand loops can be converted to integrals.
If no loop momenta vanish then the result, including the horizontal reflection, is
\begin{align}
\CT 2 4 &\supset
\frac1{2^{10}m^3L^3} \left[\frac1{L^3}\sum_{p\ne 0}-\int_p\right]
\frac{f^{\rm STT}(p^2)}{p^2}\,,
\\
f^{\rm STT}(0) &= 0\,,
\qquad\qquad
{f^{\rm STT}}'(0) = \frac{I^{\rm TT}}{32 \pi^4}\,.
\end{align}
Using Eq.~(\ref{eq:Idef}), this yields
\begin{align}
\CT 2 4 &\supset
-\frac{I^{\rm TT}}{2^{15} \pi^4 m^5L^6} 
\,.
\label{eq:STT}
\end{align}

If the left-hand loop momentum vanishes, I find
\begin{align}
\CT 2 4 &\supset 
\frac{I^{\rm TT}}{2^{14} \pi^4 m^5 L^6} 
%\,,
%\\
%I^{\rm TT} &= 8 \pi^4 \iint_{q,ik} \frac{m^3}{ \omega_k^3(\omega_k+m)\omega_q^3 (\omega_q+m)(\omega_q+\omega_k)}
%= \frac{2\pi-6}{3 }
\,,
\label{eq:STT0}
\end{align}
with $I^{\rm TT}$ given in Eq.~(\ref{eq:ITT}).

\subsubsection{SSU diagram}

Figure~\ref{fig:C2lam4}(j) combines with the $A_{2s}+A_{2u}$ contributions
on the right-hand vertices in Fig.~\ref{fig:C2lam4}(e),
the $A_{3su}$ contribution in Fig.~\ref{fig:C2lam4}(b) and the $A_{4ssu}$
counterterm in Fig.~\ref{fig:C2lam4}(a).
The fact that $A_{2u}$ contributes is not obvious but can be understood by
a careful accounting of the Wick contractions.
%I check this by the fact that UV divergences cancel.
Momentum sums in the two right-hand loops can be replaced by integrals.
I label the momenta in these loops $q$ and $k$,
while that in the left-hand loop is denoted $p$.
This is the most tedious of the diagrams to calculate. 

If $\vec p\ne 0$ then the contribution takes the form
\begin{align}
\CT 2 4 &\supset \frac1{2^8 m^5 L^3} \left[\frac1{L^3}\sum_{\vec p\ne 0}-\int_p\right]
 \frac1{p^2} \int_{q,k} g^{\rm SSU}(\vec p, \vec q, \vec k)
\,.
\end{align}
Using the fact that $g^{\rm SSU}$ vanishes when $\vec p=0$
(which again follows from the renormalization condition),
expanding $g^{\rm SSU}$ in powers of $\vec p$, 
and using Eq.~(\ref{eq:Idef}), I find
\begin{align}
\CT 2 4 
%&\supset \frac1{2^8 m^5 L^6} \iint_{q,k} h^{(j)}(\omega_q,\omega_k,\omega_{kq}) 
%\\
&= - \frac1{2^{11} \pi^4 m^5 L^6} {I^{\rm SSU}}\,,\qquad
I^{\rm SSU} = 0.156906 \,.
\label{eq:ISSU}
\end{align}

If $\vec p=0$, the result is
\begin{align}
\CT 2 4 
%&\supset \frac1{2^8 m^5 L^6} \iint_{q,k} h^{(j)}(\omega_q,\omega_k,\omega_{kq}) 
%\\
&= \frac1{2^{11} \pi^4 m^5 L^6} {I^{\rm SU}}\,,
\label{eq:SSU0}
\end{align}
where $I^{\rm SU}$ is given in Eq.~(\ref{eq:ISU}).

\subsubsection{TST diagram}

The final diagram is the boxlike Fig.~\ref{fig:C2lam4}(k), 
which is combined with the $A_{2t}+A_{2u}$ contribution to Fig.~\ref{fig:C2lam4}(e)
and the $(A_{2t}+A_{2u})^2$ contribution to Fig.~\ref{fig:C2lam4}(c).

The sums over the outer momenta can be replaced by integrals.
If the central loop momentum (denoted $\vec p$) is nonvanishing, I find
\begin{align}
\CT 2 4 &\supset
\frac1{2^{10}m^5 L^3} \left[\frac1{L^{3}} \sum_{p\ne 0} - \int_p\right]
\frac{f^{\rm TST}(p^2)}{p^2}\,,
\\
f^{\rm TST}(0)&=0\,,\qquad
f^{\rm TST'} = \frac{I^{\rm TT}}{32 \pi^4}\,.
\end{align}
Using Eq.~(\ref{eq:Idef}) then yields
\begin{align}
\CT 2 4 &\supset
- \frac{I^{\rm TT}}{2^{15}m^5 L^6} 
\,.
\label{eq:TST}
\end{align}

If $\vec p=0$, the result is
\begin{align}
\CT 2 4 &\supset
\frac{I^{\rm TT}}{2^{15} m^5 L^{6}} 
\,.
\label{eq:TST0}
\end{align}
Thus the total contribution from this diagram vanishes.

\subsubsection{Total contribution to $\Delta E_{2,\thr}^{(4)}$}

Combining the results from 
Eqs.~(\ref{eq:SSS}), (\ref{eq:SST0}), (\ref{eq:STS}), (\ref{eq:STS0}), 
(\ref{eq:STT}), (\ref{eq:STT0}), (\ref{eq:ISSU}), (\ref{eq:SSU0}), 
(\ref{eq:TST}) and (\ref{eq:TST0}), I obtain
\begin{align}
a_2^\four &\supset
-\mathcal I^3 + 6 \mathcal I \mathcal J - 3 \mathcal K 
-2^{5}\pi^2 
\left(I^{\rm SST0}- I^{\rm STS} +I^{\rm STS0}+\tfrac14 I^{\rm TT}-4 I^{\rm SSU} +4 I^{\rm SU}\right)
\,.
\label{eq:CT24tot}
\end{align}

\subsection{Mass and wave-function renormalization}
\label{sec:renorm2}

\begin{figure}[tbh]
\begin{center}
\includegraphics[scale=0.5]{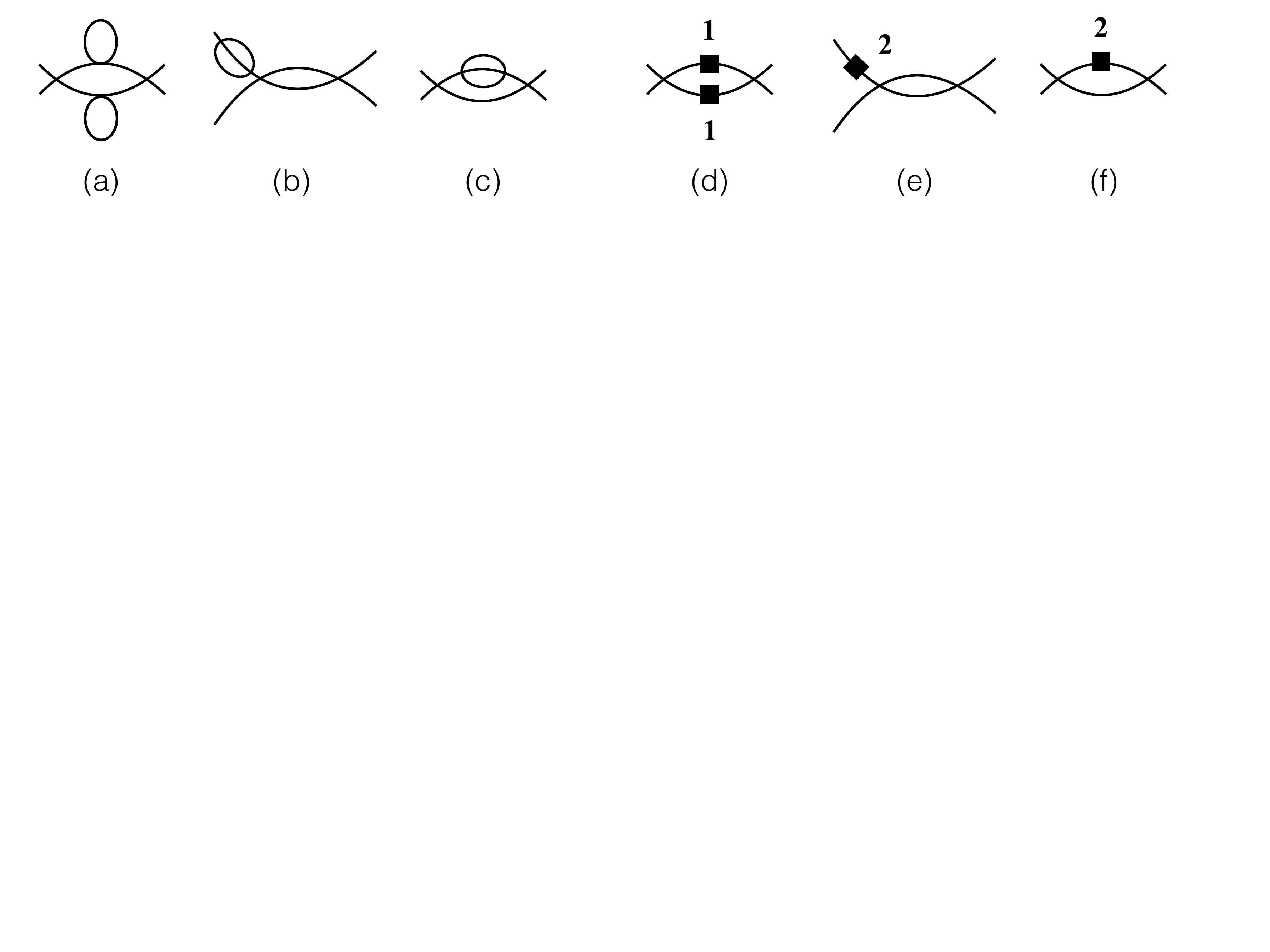}
\vskip -4truein
\caption{Examples of Feynman diagrams contributing to $\CT 2 4$
involving mass and wave-function renormalization subdiagrams.
Mass and wave-function counterterms are indicated by filled boxes.}
\label{fig:renorm}
\end{center}
\end{figure}

Both $\CZ 2 3$ and $\CT 2 4$ receive contributions from many diagrams involving
mass and wave-function renormalization parts. 
Examples are shown in Fig.~\ref{fig:renorm}.
In all cases loop sums can be replaced by integrals.
As explained in Ref.~\HSPT, tadpole bubbles, 
such as those in Fig.~\ref{fig:renorm}(a), cancel identically with the corresponding
counterterms, here shown in Fig.~\ref{fig:renorm}(d). 
For loop diagrams such as those in Figs.~\ref{fig:renorm}(b) and (c),
however, the cancelation with the counterterms of Figs.~\ref{fig:renorm}(e) and (f)
is not exact. When one constructs the renormalized propagator
by the usual geometric sum, what remains are contact terms in position space.
These, however, cannot go on shell, and thus cannot be cut, so loops involving them
do not lead to finite-volume dependence.
Instead, either they lead to contributions to the amplitudes $A_{j,k}$
[see Eq.~(\ref{eq:amps})] which thus cancel from $\Delta E_2$---exemplified
by the case of Fig.~\ref{fig:renorm}(b)---or
their contribution is canceled by coupling-constant counterterms---as is the case
for Fig.~\ref{fig:renorm}(c).
I have checked this explicitly for several examples.
The net result is that this class of diagrams does not need to be considered.

\subsection{Total result and comparison with expectation}

Combining the results in Eqs.~(\ref{eq:DE24term22}), 
(\ref{eq:DECT3}) and (\ref{eq:CT24tot}) gives the final
result for the two-particle energy shift
\begin{align}
a_2^\four &=-\cI^3 + 3 \cI \cJ - \cK
-2^{5}\pi^2 
\left(I^{\rm SST0} - 2 I^{\rm ST}- I^{\rm STS} +I^{\rm STS0}-4 I^{\rm SSU} + 2 I^{\rm SU}\right)
\,.
\label{eq:a24fin}
\end{align}
This should be compared to the result expected from the quantization condition,
Eq.~(\ref{eq:DE24}), which yields
\begin{align}
a_2^\four &= 
-\cI^3 + 3\cI \cJ - \cK
- 2^{12} \pi^6 \cK'^{(3)}_{2,s,\thr}
\,.
\label{eq:a24pred}
\end{align}
The coefficients of the geometric constants agree. For the remaining part the
result for $\cK'^{(3)}_{2,s,\thr}$ from Appendix~\ref{app:K} is needed.
Combining Eqs.~(\ref{eq:KSU}) and (\ref{eq:KST}), the result is
\begin{equation}
\cK'^{(3)}_{2,s,\thr} = \frac{I^{\rm STr}}{2^8 \pi^4} + \frac{I^{\rm SUr}}{2^6 \pi^4}
\,.
\end{equation}
Agreement between Eqs.~(\ref{eq:a24fin}) and (\ref{eq:a24pred})
holds because of the numerical relations
\begin{align}
I^{\rm SST0} - 2 I^{\rm ST} - I^{\rm STS} + I^{\rm STS0} &= \tfrac12 I^{\rm STr} \,,
\\
-4 I^{\rm SSU} + 2 I^{\rm SU} &= 2 I^{\rm SUr}\,.
\end{align}
From the point of view of the present calculation this agreement appears highly nontrivial,
as the two sides of these equations are obtained in very different ways.
I stress that the agreement holds separately for subsets of diagrams:
the SSU contribution to $\CT 2 4$, combined with the SU contribution to $\CZ 2 3 \CT 2 1$
matches with $\cK'^{(3,su)}_{2,s,\thr}$,
while the SST and STS contributions to $\CT 2 4$, combined with the ST contribution
to $\CZ 2 3\CT 2 1$, matches with $\cK'^{(3,st)}_{2,s,\thr}$.
This diagram-level matching holds also for the SSS, STT and TST classes of diagrams,
where there is no contribution to $\cK'^{(3)}_{2,s,\thr}$.

\section{Determining $\Delta_{32}^{(4)}$}
\label{sec:DE34}

In this section I calculate $\Delta_{32}^{(4)}$ in order to test the result
(\ref{eq:D32calc}) obtained from the three-particle quantization condition.
It is convenient to write
\begin{equation}
\Delta_{32}^{(4)} = \frac{3 a_3^\four}{ 2^{18} \pi^6 m^5 L^6} + \cO(L^{-7})
\end{equation}
and quote results for $a_3^\four$.
As shown in Eq.~(\ref{eq:DE324}), the calculation requires determining
$\CT 2 4$, $\CZcon 3 3$ and $\CTcon 3 4$. The former was worked out in
Sec.~\ref{sec:CT24}, and from Eq.~(\ref{eq:CT24tot}) I find
\begin{align}
a_3^\four\Big|_{\CT 2 4} &=
2\mathcal I^3 -12 \mathcal I \mathcal J +6 \mathcal K 
+
2^{6}\pi^2 
\left(I^{\rm SST0}+ I^{\rm STS0} -I^{\rm STS}+\tfrac14 I^{\rm TT}+4 I^{\rm SU}-4 I^{\rm SSU} \right)
\,.
\label{eq:a34CT24}
\end{align}
In the following two subsections I calculate the other two required quantities.

\subsection{Calculation of $\CZcon 3 3$}
\label{sec:CZcon33}

\begin{figure}[tbh]
\begin{center}
\includegraphics[scale=0.5]{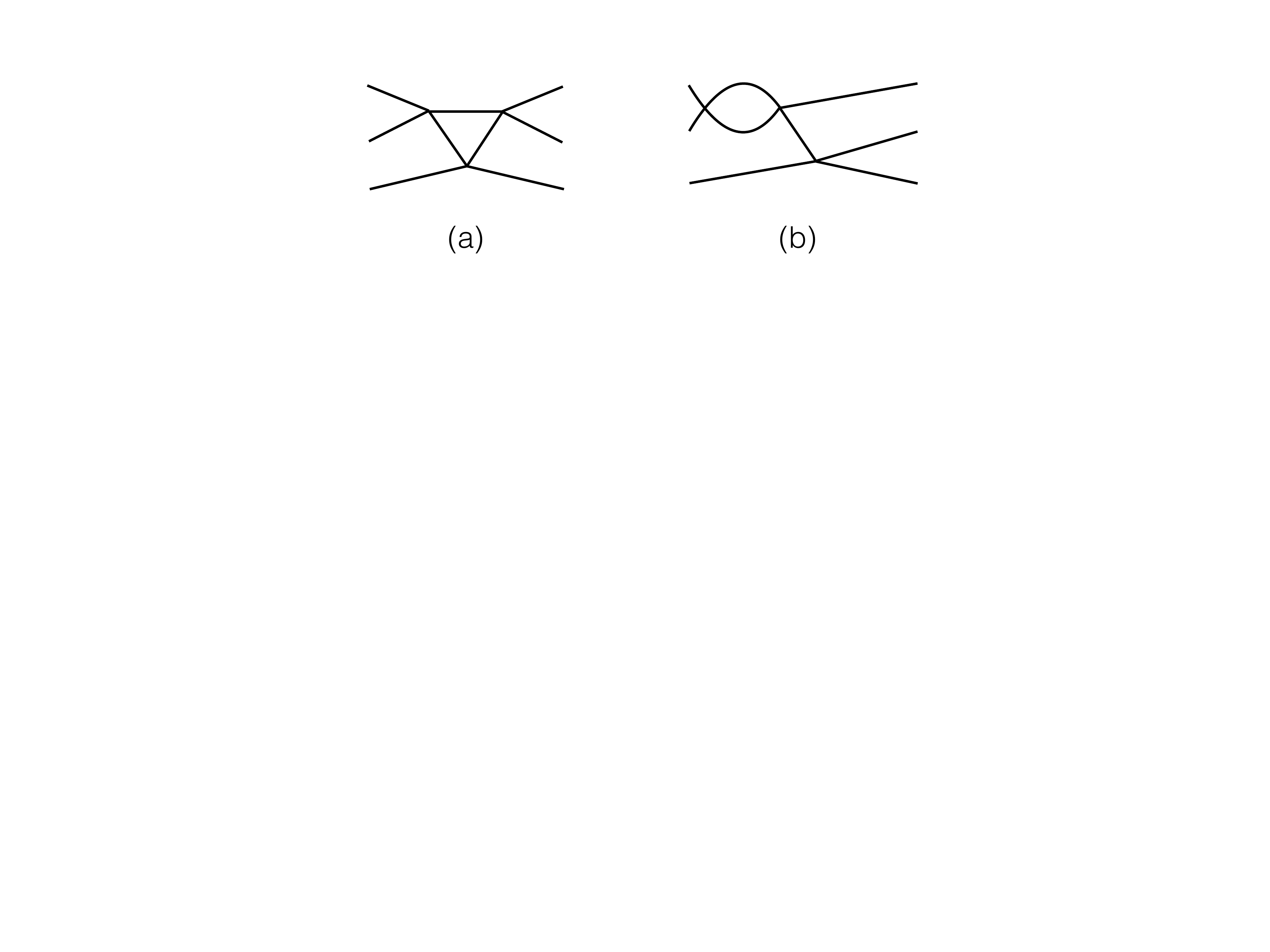}
\vskip -3.9truein
\caption{Feynman diagrams for $\CZcon 3 3$ that give contributions of $\cO(L^{-3})$.}
\label{fig:C33}
\end{center}
\end{figure}

In order to give rise to an $\cO(L^{-6})$ contribution to $\Delta_{32}^\four$,
$\CZcon 3 3$ must scale as $L^{-3}$.
Since connected diagrams begin at $\cO(1/L^6)$, to reach the required dependence
requires a $1/p^6$ IR divergence. This is possible in third order diagrams
only if there is a time ordering of vertices in which all intermediate states contain only
three particles (for each such intermediate state can yield a factor of $1/p^2$).
This singles out the two diagrams shown in Fig.~\ref{fig:C33},
which I denote, following Ref.~\HSPT, as (a)
the bull's head and (b) the s-channel fish diagram.

The calculation is very straightforward as only one time ordering is required.
The contribution from the bull's head diagram is
\begin{align}
\CZcon 3 3 &\supset 
- \frac{3}{2^7 m^3 L^9} \sum_{\vec p\ne 0} \frac{-2 + \cO(p^2)}{p^6} 
= \frac{3}{2^{13} \pi^6 m^3 L^3} (2 \cK) + \cO(1/L^4)
\,,
\end{align}
while that from the s-channel fish (together with its horizontal reflection) is
\begin{align}
\CZcon 3 3 &\supset 
- \frac{3}{2^7 m^3 L^9} \sum_{\vec q\ne 0} \frac{1+ \cO(q^2)}{q^6} 
= \frac{3}{2^{13} \pi^6 m^3 L^3} (-\cK) + \cO(1/L^4)
\,.
\end{align}
In both cases I have used Eq.~(\ref{eq:Kdef}).
Combining these results and multiplying by $3\,\CT 2 1$ yields
\begin{equation}
a_3^\four\Big|_{\CZcon 3 3} = -12 \cK
\,.
\label{eq:a34CZ33}
\end{equation}

\subsection{Calculation of $\CTcon 3 4$}
\label{sec:CTcon34}

\begin{figure}[tbh]
\begin{center}
\includegraphics[scale=0.5]{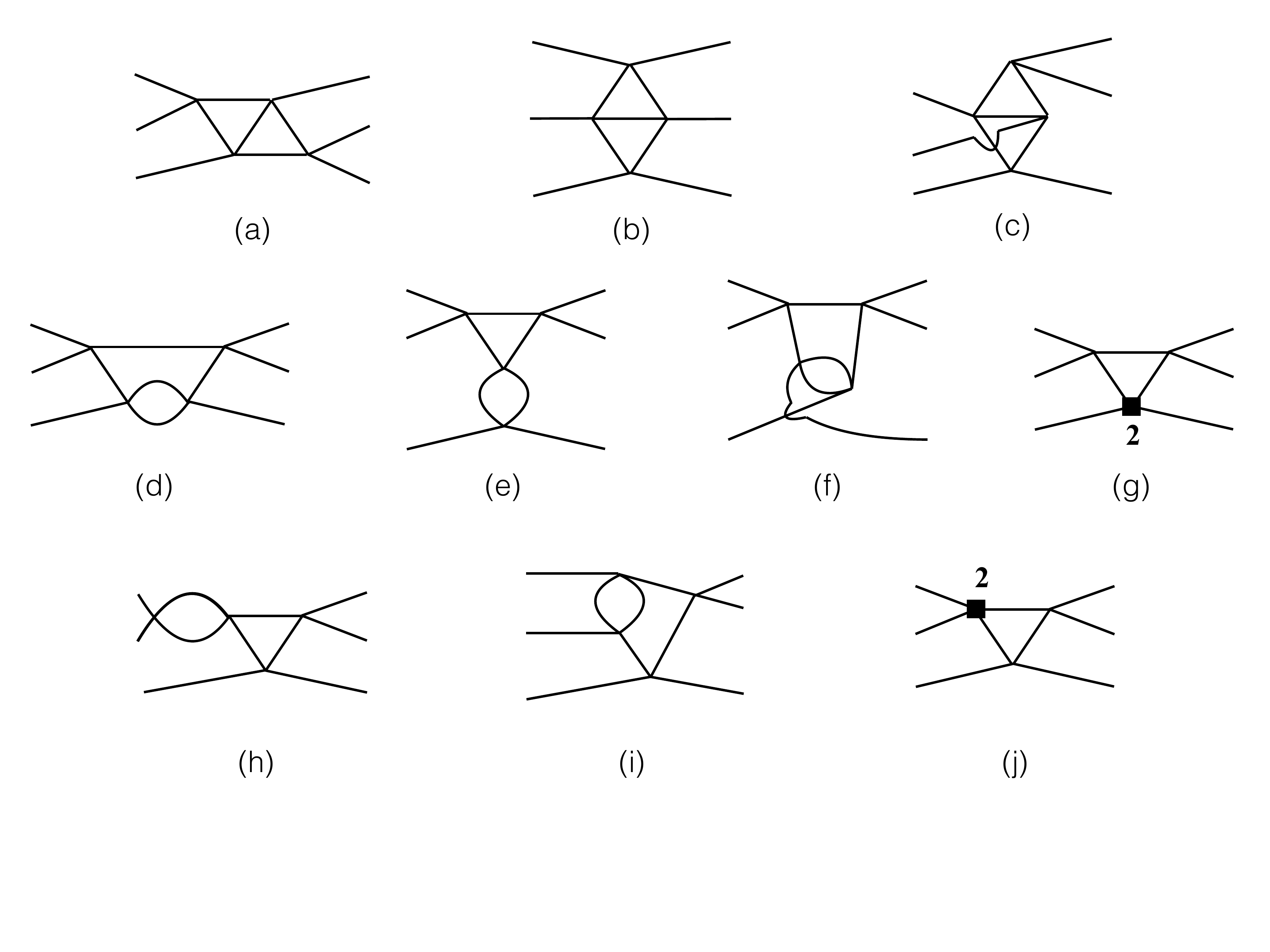}
\vskip -0.7truein
\caption{A subset of Feynman diagrams contributing to $\CT 3 4$ and $\Mthr$.
Solid squares are vertex counterterms, with the number
indicating the power of $\lambda$.
External particles have zero three-momentum. 
Diagrams related by vertical or horizontal reflection are not shown.}
\label{fig:3part1}
\end{center}
\end{figure}

At $\cO(\lambda^4)$, connected three-particle diagrams contain two loops.
A selection of the many such diagrams is shown in
Figs.~\ref{fig:3part1}, \ref{fig:3part2} and \ref{fig:3part3}, including the
subset that will need to be considered in detail.
As for $\CT 2 4$, the diagrams have an initial volume scaling of $L^{-12}$.
This can be raised to the desired $L^{-6}$ dependence
either by converting two sums over intermediate momenta to integrals
or by having a single loop sum that diverges in the IR as $1/p^6$.
The latter case requires that the second loop momentum vanishes.\footnote{%
This is the most IR singular summand possible at this order
because, of the four integrals over the times of the vertices,
one is needed to produce the factor of $\tau$, while each of the other three
can lead to a factor of $1/p^2$.}
As already noted above, to obtain the most singular IR divergence the diagram
must be such that there is a time ordering in which all intermediate states involve
three particles. The diagrams for which this holds are Figs.~\ref{fig:3part1}(a), (d) and (g),
Fig.~\ref{fig:3part2}(a) and Fig.~\ref{fig:3part3}(a).

If both loop momenta vanish then the contributions are proportional to $1/L^{12}$ and
thus of too high order.

When both loop momenta are nonzero, there are two further general results
that simplify the calculations.
The first is that, as for $\CT 2 4$,  the vertex times, $\tau_i$, must satisfy $0 < \tau_i < \tau$.
%This result follows from the fact that the $\tau_i$ must lie close to each other
%in order to avoid an exponential suppression. Thus the vertices are ``clumped''.
%The only way to obtain a term linear in $\tau$ (after removal of excited-state contributions) 
%is then for the clump of vertices to be integrated from $0$ to $\tau$
%(with the exponential factors from the initial and final propagators being independent
%of the position of the clump). This leads to the condition quoted above.
The second concerns the summand
that remains after the time integrals are done (leaving only momentum sums). 
For $\CT 2 4$, this summand was proportional to the integrand of $\cM_{2}$ at threshold.
Here, by a similar argument, one can show that the summand
of a contribution to $\CTcon 3 4$, when multiplied by  $48 m^3 L^{12}$,
gives the integrand of $\cM_3$ at threshold.\footnote{%
The factor of $48$ can be understood from the case of a local
$\lambda_6\phi^6/6!$ interaction, for which $\cM_3=-\lambda_6$.
The contribution to $\CTcon 3 4$ is then $-\lambda_6/[6 (2m)^3]$,
with the $6$ arising from the numerator in ratio defining $\CT 3 4$,
Eq.~(\ref{eq:C3def}), and the $(2m)^{-3}$
arising from the three propagators that are not canceled in this ratio.}
This implies that if both loop sums can be replaced by integrals, which is allowed
in the absence of IR divergences,
then the diagram will give a contribution of the form\footnote{%
In the remainder of this section the fact that there
are contributions of $\cO(L^{-7})$ will not be noted explicitly.}
\begin{equation}
\CTcon 3 4 \supset \frac{\Mthr}{48 m^3 L^6} + \cO(L^{-7})
\,.
\label{eq:standard}
\end{equation}
I will refer to this as the ``standard form" of contribution.

Thus the only diagrams that need to be considered in detail are those
containing IR divergences. These arise when the diagram has three-particle cuts.
Thus, for example, Fig.~\ref{fig:3part1}(b) need not be considered, since it
has no three-particle cuts and thus contributes only to the standard form,
Eq.~(\ref{eq:standard}). 
All diagrams having three-particles cuts are included in
Figs.~\ref{fig:3part1}, \ref{fig:3part2} and \ref{fig:3part3},
with the exception of those with self-energy insertions or that are one-particle
reducible. The latter do not lead to nonstandard contributions and are discussed
in Sec.~\ref{sec:remaining}.

A further distinction allows a subset of the diagrams with three-particle cuts to be
removed from consideration. If the IR divergence occurs inside a loop, then
it must be stronger than $1/p^2$ in order for Eq.~(\ref{eq:standard}) to be invalidated,
as such an IR divergence is integrable.
Since each three-particle cut only leads to a $1/p^2$ divergence, 
this implies that, for diagrams in which the three-particle cuts run through loops,
there must be at least two such cuts in order to
obtain a result different from Eq.~(\ref{eq:standard}).
Thus Fig.~\ref{fig:3part1}(c) need not be considered.
The alternative is that  the single three-particle cut does not pass through a loop, 
which is the case for Fig.~\ref{fig:3part2}(c) and
Figs.~\ref{fig:3part2}(d), (e) and (f). These diagrams can lead to contributions
of a form differing from Eq.~(\ref{eq:standard}) and must be considered in detail.

The final general issue arises from the fact that $\cM_3$ at threshold is IR divergent
and thus ill-defined. This is not the case for $\cM_2$ and adds another level
of complication to the three-particle analysis. To obtain a well-defined
three-particle amplitude at threshold one must add an IR regulator, make
some subtractions, and then remove the regulator~\HSQCa.
The choice of subtraction introduces scheme dependence, and a particularly
simple choice was introduced in Ref.~\HSTH\ and used to define the
quantity $\Mthr$ that occurs in the prediction that I am testing, Eq.~(\ref{eq:DE324}).
The general implication is that, for diagrams with IR divergences that are not integrable,
one must determine both their contributions to $\CTcon 3 4$ and, in a separate
infinite-volume calculation, to $\Mthr$, so that the
deviation from the standard result (\ref{eq:standard}) can be found.
In the following I work systematically through all such diagrams carrying out this procedure.

It will be useful to have in mind the form of the IR subtractions that are needed.
These are defined in Eq.~(114) of Ref.~\HSTH\ and the subsequent text.
The schematic form is 
\begin{equation}
\Mthr \equiv \lim_{\delta\to 0} \left\{
\cM_{3,\delta} - I_{0,\delta} - \int_{k_1,\delta} \Xi_1(\vec k_1)
- \int_{k_1,\delta} \int_{k_2,\delta} \Xi_2(\vec k_1, \vec k_2) \right\}
\,.
\label{eq:Mthrdef}
\end{equation}
Here $\delta$ is an IR regulator defined such that threshold is attained when $\delta\to 0$.
The specific form of this regulator, as well as the explicit expressions for $I_{0,\delta}$,
$\Xi_1$ and $\Xi_2$, will be given below when needed.
$I_{0,\delta}$ contains terms of all orders in $\lambda$ starting at $\lambda^2$, 
while $\Xi_1$ contains terms proportional to $\lambda^3$ and $\lambda^4$,
and $\Xi_2$ is proportional to $\lambda^4$.
Thus several new features of the subtraction scheme are being tested by working at $\cO(\lambda^4)$.
I also note that $I_{0,\delta}$ is used to subtract IR divergences in the diagrams of
Figs.~\ref{fig:3part2} and \ref{fig:3part3}, while $\Xi_1$ and $\Xi_2$ are needed for some of
the diagrams in Fig.~\ref{fig:3part1}.

\subsubsection{Double triangle diagram: Fig.~\ref{fig:3part1}(a)}

This is the first example of a class of diagrams arising first at fourth order,
which involve a double triangle or diamond.
Its IR behavior arises from a process in which there are three $2\to 2$ scatterings, with
the spectator particle alternating.
As shown in Ref.~\cite{Hansen:2016fzj}, this leads to a logarithmic IR divergence in the
corresponding threshold amplitude, requiring the subtraction of
the fourth-order term $\Xi_2$. 
%The diagram is UV finite.

If one of the loop momenta vanishes, the result is
\begin{align}
\CTcon 3 4 &\supset \frac{3}{2^8 m^5 L^{12} }
\sum_{\vec p \ne 0} \frac{-2 + \cO(p^2)}{p^6}
\ \ \Rightarrow\ \
a_3^\four \supset  32 \cK
\,,
\label{eq:res1}
\end{align}
where I have used Eq.~(\ref{eq:Kdef}) to obtain the second result.

The result if both loop momenta are nonzero is
\begin{align}
\CTcon 3 4 &\supset
\frac{1}{48 m^3 L^{12}} 
\sum_{\vec p\ne 0}\sum_{\vec q\ne 0} 
\frac{9}{16 } \frac{g(\vec p,\vec q)}{p^2 q^2 (W_{pq}^2 - 9 m^2)}\,,
\label{eq:DTresa}
\end{align}
where $W_{pq}=\omega_p+\omega_q + \omega_{pq}$,
$\omega_{pq}^2 =m^2+ (\vec p+\vec q)^2$, and $g(\vec p,\vec q)$ 
is a nonsingular function that I do not reproduce, except to note
that $g(0,0)=3/m^2$. The logarithmic IR divergence can be seen by noting
that, for small momenta,
\begin{equation}
W_{pq}^2 - 9 m^2 = 6\left(p^2 + q^2 + \vec p\cdot\vec q \right) + \cdots\,.
\label{eq:Wpqexp}
\end{equation}

As explained in the introduction to this section, the summand in Eq.~(\ref{eq:DTresa})
is the integrand of the contribution of the double-triangle diagram to $\cM_3$ at threshold. 
The prescription of Ref.~\HSTH\ to remove the IR divergences in this case is to subtract
the quantity
\begin{equation}
\frac1{\lambda^4}\Xi_2(\vec p,\vec q) = \frac{9 }{16 m^2}
\frac{H(\vec p)^2 H(\vec q)^2}{p^2 q^2 [p^2 + q^2 + (\vec p+\vec q)^2]}
\,,
\end{equation}
where $H(\vec p)$ is a UV regulator whose detailed form will not matter here other than
the property $H(0)=1$.
After subtraction the result can be integrated and defines the contribution of this
diagram to the threshold amplitude, which I label $\Mthr^{\rm DT}$.
Thus I proceed by adding and subtracting the $\Xi_2$ term, leading to
\begin{align}
\CTcon 3 4 &\supset
\frac{1}{48 m^3 L^{12}} 
\sum_{\vec p\ne 0}\sum_{\vec q\ne 0} 
\left[\frac{9}{16} \frac{g(\vec p,\vec q)}{p^2 q^2 (W_{pq}^2 - 9 m^2)} 
- \frac{\Xi_2(\vec p,\vec q)}{\lambda^4}\right]
+
\frac{1}{48 m^3 L^{12}} 
\sum_{\vec p\ne 0}\sum_{\vec q\ne 0} \frac{\Xi_2(\vec p,\vec q)}{\lambda^4}
\\
&=
\frac{1}{48 m^3 L^{6}} 
\int_{p,q}
\left[\frac{9}{16} \frac{g(\vec p,\vec q)}{p^2 q^2 (W_{pq}^2 - 9 m^2)} 
- \frac{\Xi_2(\vec p,\vec q)}{\lambda^4}\right]
+
\frac{1}{48 m^3 L^{12}} 
\sum_{\vec p\ne 0}\sum_{\vec q\ne 0} \frac{\Xi_2(\vec p,\vec q)}{\lambda^4}
%+ \cO(L^{-7})
\\
&= \frac{\Mthr^{\rm (4,DT)}}{48 m^3 L^{6}} 
+
%\frac{3}{2^{18} \pi^6 m^5 L^{6}} \left( \frac{64\pi^4}3 \log N_{\rm cut} - {\cC_5}\right)
%\frac{12\pi^2}{m L^6} \left(\frac1{32 \pi m}\right)^4 \left( \frac{64\pi^4}3 \log N_{\rm cut} - {\cC_5}\right)
\frac{3}{2^{18} \pi^6 m^5 L^6}
\left( \frac{64\pi^4}3 \log N_{\rm cut} - {\cC_5}\right)
%+ \cO(L^{-7})
\,.
\label{eq:DTresc}
\end{align}
In the second line, the sum of the IR regulated difference has been replaced by an integral,
which is valid up to  a $1/L^7$ contribution arising from the 
difference between the sum and integral of an integrand with a $1/p^2$ divergence.
To obtain the final form, the expression for the sum over $\Xi_2$ given in
Eqs.~(C18) and (C19) of Ref.~\cite{Hansen:2016fzj} has been used.

\subsubsection{Diver diagram: Fig.~\ref{fig:3part1}(d)}

This diagram is combined with the $A_{2s}$ part of the counterterm diagram
Fig.~\ref{fig:3part1}(f). 
This turns out to be the most involved calculation from Fig.~\ref{fig:3part1}.
I denote the momentum in the outer loop by $p$, while that in the
diver's head loop is denoted by $q$.

If $\vec p=0$, the IR divergence is sufficient to lead to a contribution at the
desired order, specifically
\begin{align}
\CTcon 3 4 &\supset \frac{3 }{2^{10} m^5 L^6}
\frac1{L^6} \sum_{\vec q\ne 0} \frac{1 + \cO(q^2)}{q^6}
\ \ \Rightarrow\ \
a_3^\four \supset -4 \cK
\,.
\label{eq:res2}
\end{align}

If $\vec p \ne 0$, Fig.~\ref{fig:3part1}(d) alone gives
\begin{align}
\CTcon 3 4 &\supset
\frac{3}{2^{11} m^3 L^{12}} \sum_{\vec p\ne 0} \sum_{\vec q}
\frac1{\omega_p^3\omega_q}
\frac{g^D(\vec p, \vec q)}{p^4 (W_{pq}^2-9m^2)}
\,,
\label{eq:diver1}
\end{align}
where $g^D$ is a complicated function 
that is finite when $p$ and/or $q$ vanish.
Thus the summand does not diverge when $\vec q=0$, 
allowing the sum over $\vec q$ to include this point.
The sum over $\vec q$ is UV divergent, but this is canceled by the
counterterm contribution, which is
\begin{align}
\CTcon 3 4 &\supset
- \frac{3 A_{2s}}{2^8 m^3 L^9} \sum_{\vec p\ne 0}
\frac{3\omega_p^2-m^2}{\omega_p^3 p^4}
\,.
\end{align}
Combining, I find 
\begin{align}
\CTcon 3 4 &\supset
\frac{3}{2^{11} m^3 L^{6}} \left( S_1 + S_2\right)
\,,
\label{eq:S1S2}
\\
S_1 &=
\frac1{L^3}\sum_{\vec p \ne 0}  \frac{f^D(p)}{\omega_p^3 p^4} 
\,,
\\
f^D(p) &=  \int_q \frac1{\omega_q q^2} 
\left[ {q^2} \frac{g^D(\vec p, \vec q)}{(W_{pq}^2-9m^2)}
- (3 \omega_p^2-m^2) \right]
\,,
\\
S_2 &= \frac1{L^3}\sum_{\vec p \ne 0} \frac1{\omega_p^3 p^4}
 \left[\frac1{L^3}\sum_{\vec q} - \int_q\right]
\frac{g^D(\vec p, \vec q)}{\omega_q (W_{pq}^2-9m^2)}
\,.
\label{eq:S2def}
\end{align}

Consider first $S_1$.
Its IR behavior is determined by the form of $f^D(p)$ near $p=0$. 
Using the explicit form of $g^D$ I find that $f^D(0)=0$.
To next pull out the leading IR behavior of
the integrand using Eq.~(\ref{eq:Wpqexp}):
\begin{align}
f^D(p) 
&= \tilde f^D(p) + 2m I^D(p) \,,
\\
\tilde f^D(p) &= \int_q\left[\frac{g^D(\omega_p,\omega_q,\omega_{pq})}{\omega_q(W^2-9m^2)}
-\frac{2m}{p^2+q^2+\vec p\cdot\vec q} -\frac{3\omega_p^2-m^2-2m\omega_q}{\omega_q q^2}\right]\,,
\\
I^D(p) &= \int_q \left[ \frac1{p^2 + q^2 + \vec p\cdot \vec q} - \frac1{q^2}\right]\,,
\end{align}
The key property of the residue function is that $f^D(p) \propto p^2$ near $p=0$.
The integral $I^D$ is well defined as long as one does the angular integral first,
and gives
\begin{align}
I^D(p) 
%&= \frac1{4\pi^2} \int_0^\infty dq \int_{-1}^1 dc
%\left[ \frac{q^2}{p^2 + q^2 + p q c} - 1\right]
&= - \frac{\sqrt{3}\, p}{8 \pi}
\,.
\end{align}
This shows that $f^D(p)$ is a function of $p$ and not $p^2$.
Combining these results yields the $S_1$ contribution to $\CTcon 3 4$:
\begin{align}
\CTcon 3 4 &\supset
\frac{1}{48 m^3 L^6} \frac{9}{2^7 L^{3}} \sum_{\vec p\ne 0} 
\left[- \frac{\sqrt3 m }{4 \pi \omega_p^3 p^3} 
+ \frac{\tilde f^D(p)}{\omega_p^3 p^4} \right]
\,.
\label{eq:diverfinal}
\end{align}

The next step is to express this result in terms of the
contribution of the diver diagram to $\Mthr$, which I denote $\Mthr^{\rm D}$,
and determine the remainder.
Using the general result described in the introduction to this subsection,
it follows from Eq.~(\ref{eq:diverfinal}) that the contribution of the diver diagram
to the amplitude at threshold is 
\begin{equation}
\mathcal M_3^{\rm (4,D)} =  \frac{9}{2^7} \int_p
\left[- \frac{\sqrt3 m }{4 \pi \omega_p^3 p^3} 
+ \frac{\tilde f^D(p)}{\omega_p^3 p^4} \right]\,.
\end{equation}
This is IR divergent, and to obtain $\Mthr^{\rm (4,D)}$ 
one must subtract from this integral of the $\lambda^4$ part of $\Xi_1$.
The full expression for $\Xi_1$ is [see Eqs.~(191)-(121) of Ref.~\HSTH]
\begin{equation}
\Xi_1(p) = -  \frac{9\lambda^3}{8m}\left[
\frac{ H(\vec p)^2}{p^4} + \frac{\lambda}{32\pi m} \frac{\sqrt3}{2} \frac{H(\vec p)^3}{p^3} \right]
\,,
\label{eq:Xi1def}
\end{equation}
and I need here only the second term.
Thus I find
\begin{equation}
\Mthr^{\rm (4,D)} =  \frac{9}{2^7} \int_p
\left[- \frac{\sqrt3 m }{4 \pi \omega_p^3 p^3} + \frac{\sqrt3 H(\vec p)^3}{4 \pi m^2 p^3}
+ \frac{\tilde f^D(p)}{\omega_p^3 p^4} \right]\,,
\end{equation}
which indeed is IR (as well as UV) convergent.
This allows the result (\ref{eq:diverfinal}) to be rewritten as
\begin{align}
\CTcon 3 4 &\supset
\frac{\Mthr^{\rm (4,D)}}{48 m^3 L^6}
-
\frac{3}{2^{11} m^3 L^{9}} \sum_{\vec p\ne 0} 
 \frac{\sqrt3 H(\vec p)^3}{4 \pi m^2 p^3} 
\label{eq:diverfinal2}
%\\
%&=
%\frac{\Mthr^{\rm D}}{48 m^3 L^6}
%-
%\frac{3\sqrt3 }{2^{16} \pi^4 m^5 L^{6}} \sum_{\vec n_p\ne 0} 
%\frac{ H(2\pi\vec n_p/L)^3}{n_p^3} 
\\
&=
\frac{\Mthr^{\rm (4,D)}}{48 m^3 L^6}
-
\frac1{48 m^3 L^6}\frac{\chi_{1,B}}{\lambda^4}
\\
&=
\frac{\Mthr^{\rm (4,D) }}{48 m^3 L^6}
-
\frac{3}{2^{18} \pi^6 m^5 L^6}
\left( 16 \pi^3 \sqrt3 \log N_{\rm cut} + {\cC_4}\right)
\,.
\label{eq:res3}
\end{align}
In the second step I use the definition of
$\chi_{1,B}$ given in Eq.~(C13) of Ref.~\cite{Hansen:2016fzj}, 
and in the last step I use the evaluation of $\chi_{1,B}$ presented in
Eq.~(C15) of that work.

\bigskip
Now I turn to $S_2$. Naively, it appears that the sum-integral difference appearing in
Eq.~(\ref{eq:S2def}) is exponentially suppressed, because there are no singularities
in the region of integration over $\vec q$ when $\vec p\ne 0$ (since $W_{pq}> 3m$).
However the singularities are nearby, 
and the ``suppression" is by $\exp(-p L)\sim \cO(L^0)$.
It follows that this term must be kept. 
Fortunately, it turns out that it can be related analytically to
the $\cC_F$ contribution to $\Delta_{32}^\four$ in Eq.~(\ref{eq:DE324}).

To do so I rewrite $S_2$ by setting $W_{pq}\to 3m$ everywhere
except for the $1/(W_{pq}-3m)$ pole. This leads only to
corrections suppressed by $\exp(-mL)$.
Using the result
\begin{equation}
\frac{g^D(\omega_p,\omega_q,\omega_{pq})}{W_{pq}+3m}\Bigg|_{W_{pq}=3m}
= \frac{\omega_p^2 (m+\omega_p)^2}{2m^2}
\,,
\end{equation}
I find
\begin{align}
S_2 &= -  \frac1{L^3}\sum_{\vec p \ne 0} \frac1{p^4} \frac{(m+\omega_p)^2}{2m^2}
 \left[\frac1{L^3}\sum_{\vec q} - \int_q\right]
\frac{1}{\omega_p \omega_q \omega_{pq} (3m -W_{pq})}
\,.
\end{align}
Observing that the sum over $\vec p$ is dominated by $p \sim 1/L$, 
and dropping higher order corrections in $1/L$, this can be rewritten as
\begin{align}
S_2 &= -  \frac1{L^3}\sum_{\vec p \ne 0} \frac2{p^4} 
 \left[\frac1{L^3}\sum_{\vec q} - \int_q\right]
\frac{1}{\omega_p \omega_q \omega_{pq} (3m -W_{pq})}
= - \frac{2^7}{9} \frac{\chi_F}{\lambda^4}\,.
\end{align}
Here $\chi_F$ is a quantity introduced in Ref.\HSTH,
which evaluates to
\begin{equation}
\chi_F 
%= \frac{9 \times 2^6 m^2 a ^4}{\pi^2} \cC_F
= \lambda^4  \frac{ 9}{2^{14} \pi^6 m^2} \cC_F
\,.
\end{equation} 
Combining these results with Eq.~(\ref{eq:S1S2}) I find the contribution of the $S_2$ term to be
\begin{align}
\CTcon 3 4 &\supset - \frac3{2^{18} \pi^6 m^5 L^6} \cC_F
\,.
\label{eq:res4}
\end{align}

\subsubsection{Figure~\ref{fig:3part1}(e)}

This diagram is combined with the
$A_{2t}$ contribution from Fig.~\ref{fig:3part1}(g).
I denote the momentum in the bull's head loop by $\vec p$ and the other
loop momentum by $\vec q$.
The sum over $\vec q$ can be replaced by an integral since the summand is IR finite.
For any nonzero choice of $\vec p$, the factorization of the two loops then implies that
there is an exact cancellation between Figs.~\ref{fig:3part1}(e) and (f).
For $\vec p=0$ the absence of an IR divergence in $\vec q$ implies that the contribution
is of $\cO(L^{-9})$. Thus these diagrams make no contribution to $\CTcon 3 4$.

They also make no contribution to $\Mthr$. 
To understand this first note that both the relevant IR subtraction terms in the definition
of $\Mthr$, Eq.~(\ref{eq:Mthrdef}), namely $\Xi_1$ and $\Xi_2$,
have already been used for the earlier diagrams.
Thus the contribution of Fig.~\ref{fig:3part1}(e) together with the counterterm must be
IR finite by itself.
This is a somewhat subtle issue since the bull's head loop alone has a
nonintegrable $1/p^4$ dependence in the IR~\HSPT.
To understand this issue requires using the IR regularization defined in Ref.~\HSTH:
external momenta are set to zero,
and an IR cutoff is applied to the loop momentum, $p \ge \delta$. 
The result should then be IR finite when $\delta\to 0$.
The point is that, since the two loops factorize, the cancelation with the counterterm
is exact for any nonzero $\vec p$, and so the contribution to $\Mthr$ vanishes for all
nonzero $\delta$ and thus also in the limit $\delta\to 0$.

\subsubsection{Figure~\ref{fig:3part1}(f)}

This diagram is combined with the
$A_{2u}$ contribution from Fig.~\ref{fig:3part1}(g).
I denote the momentum in the upper loop by $\vec p$ and that in the lower loop
by $\vec q$.
In contrast to Fig.~\ref{fig:3part1}(e), here the loops do not factorize,
because the momentum $\vec p$ passes through the lower loop.
This implies, as explained below, that the diagram (plus counterterm) contributes to
both $\CTcon 3 4$ and $\Mthr$. However, this contribution turns out to
be only of the standard form, Eq.~(\ref{eq:standard}),
so no explicit calculation is needed.

The analysis of the diagram starts by noting that, as for Fig.~\ref{fig:3part1}(e),
the sum over $\vec q$ can be replaced by an integral since the summand is IR finite,
The addition of the counterterm renders the integral finite in the UV, and evaluating
the integral leads to a function $J(p^2)$ that vanishes when $\vec p=0$ 
(since then, as for Fig.~\ref{fig:3part1}(e), the cancelation with the counterterm is exact). 
Thus $J(p^2) \propto p^2$,
and this reduces the IR divergence from $1/p^4$  to $1/p^2$.
Since the latter form is integrable, the sum over $\vec p$ can be replaced by an integral
up to $\mathcal O(1/L^7)$ corrections.
Thus, at the order I work, no finite-volume contributions arise aside from the
standard contribution involving $\Mthr$.

\subsubsection{Figure~\ref{fig:3part1}(h)}

Figure.~\ref{fig:3part1}(h) combines with the $A_{2s}$ part of 
Fig.~\ref{fig:3part1}(j), together with horizontal reflections.
Viewed as a contribution to $\Mthr$, the argument given for the previous diagram
continues to hold: there is an exact cancelation
between the s-channel loop and its counterterm.
This is not the case, however, when the diagram is evaluated as a contribution to
$\CT 3 4$. This is because the s-channel loop momentum is summed in
Fig.~\ref{fig:3part1}(h) but integrated (in $A_{2s}$) in Fig.~\ref{fig:3part1}(j).
The sum-integral difference leads to a finite-volume residue that, combined with 
the IR divergence from the bull's head diagram, leads to a $1/L^6$ correction.

There are three contributions of this type. The first occurs when both loop
momenta are nonzero:
\begin{align}
\CTcon 3 4 &\supset \frac{3}{2^9 m^5 L^6}
\frac1{L^3} \sum_{\vec p \ne 0} \frac{m^4}{\omega_p^3 p^4 }
\left[\frac1{L^3} \sum_{\vec q  \ne 0} - \int_q\right]
\frac{1 + 3 \vec p^2/(2m^2)}{\omega_q q^2}
%\\
%&= \frac{3}{2^9 m^5 L^6} \frac{\cI \cJ}{(2\pi)^6} + \cO(1/L^7)
\\
&= \frac{3}{2^{18} m^5 L^6} (8 \cI \cJ) %+ \cO(1/L^7)
\,.
\end{align}
The second arises when the s-channel loop momentum vanishes:
\begin{align}
\CTcon 3 4 &\supset \frac{3}{2^9 m^5 L^6}
\frac1{L^6} \sum_{\vec p\ne 0} \frac{-2 + \cO(p^2)}{p^6}
= \frac{3}{2^{18} \pi^6 m^5 L^6} (-16 \cK) % + \cO(1/L^7)
\,.
\end{align}
The final contribution occurs when the bull's head loop momentum vanishes:
\begin{align}
\CTcon 3 4 &\supset \frac{3 }{2^9 m^5 L^6}
\frac1{L^6} \sum_{\vec q\ne 0} \frac{1 + \cO(q^2)}{q^6}
= \frac{3}{2^{18} \pi^6 m^5 L^6} (8 \cK) % + \cO(1/L^7)
\,.
\end{align}
In total, this diagram gives
\begin{equation}
a_2^\four \supset -8\cI\cJ + 8 \cK\,.
\label{eq:res5}
\end{equation}

\subsubsection{Figure~\ref{fig:3part1}(i)}

The final diagram of this class is Fig.~\ref{fig:3part1}(i),
which combines with the $A_{2t}+A_{2u}$ part of Fig.~\ref{fig:3part1}(j).
Here the argumentation is not so straightforward since the two loops do not
factorize.
Thus, while the UV divergence is canceled by the counterterm,
there will be a finite residue. This residue vanishes, however, when the momentum
in the bull's head loop itself vanishes. This in turn implies that the IR divergence in the
bull's head loop is canceled. It then follows that the difference between the
momentum sum and integral is exponentially suppressed, so that the contribution
to $\CT 3 4$ is simply of the standard form, Eq.~(\ref{eq:standard}).
%\begin{equation}
%\CTcon 3 4 \supset \frac{\Mthr^{\rm (h)}}{48 m^3 L^6}
%\,.
%\label{eq:res6}
%\end{equation}

\begin{figure}[tbh]
\begin{center}
\includegraphics[scale=0.5]{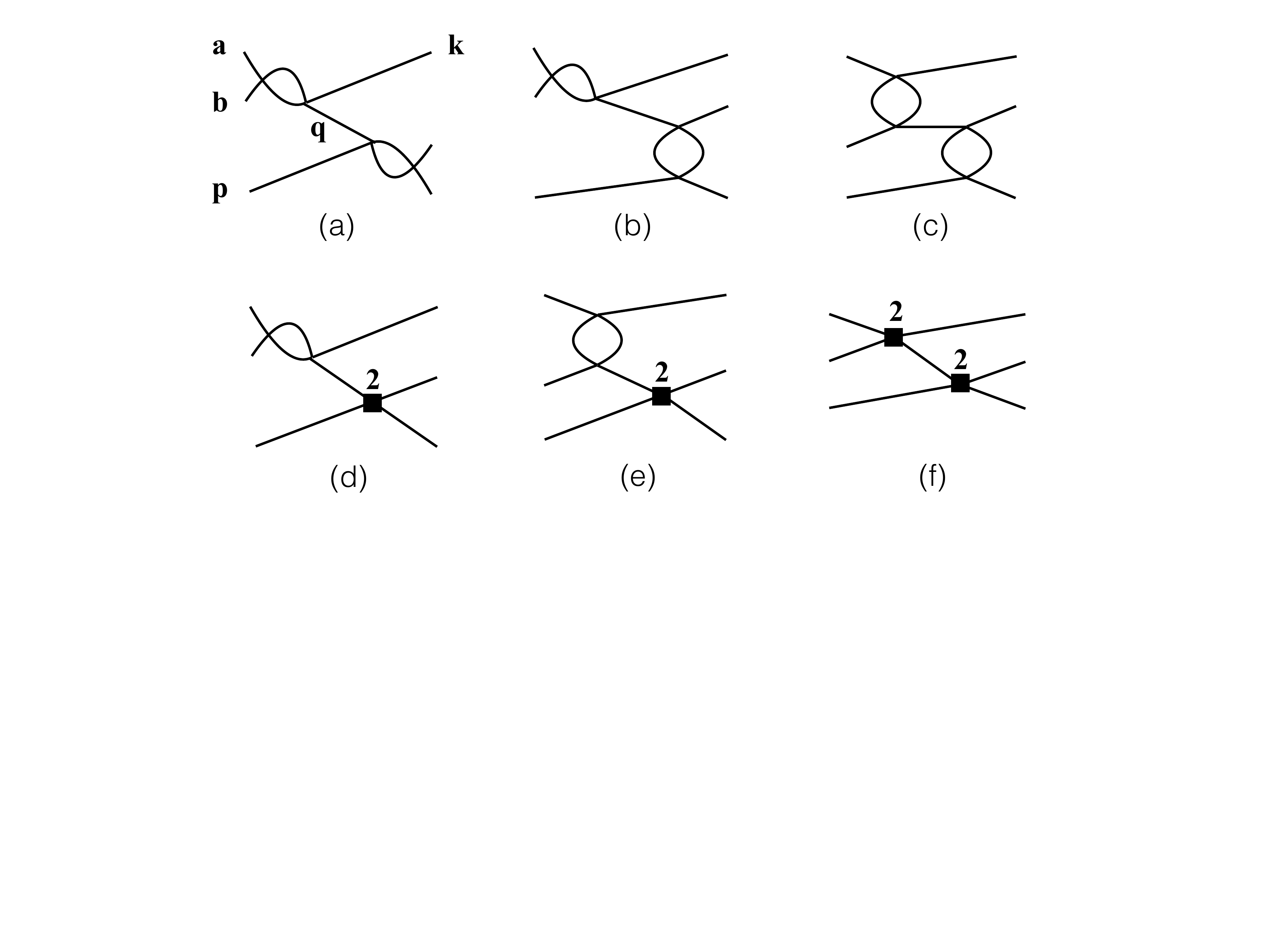}
\vskip -2.4truein
\caption{Further diagrams contributing to $\CTcon 3 4$ and $\Mthr$. Reflections are not shown.}
\label{fig:3part2}
\end{center}
\end{figure}

\subsubsection{Double fish diagram: Fig.~\ref{fig:3part2}(a)}
\label{sec:2Sfish}

I now turn to the two-loop radiative corrections to the three-particle
tree diagram, starting with those involving two separate loops,
shown in Fig.~\ref{fig:3part2}.
The first diagram is that containing two s-channel fish,
Fig.~\ref{fig:3part2}(a), which combines with the contributions to
Figs.~\ref{fig:3part2}(d) and (f) in which the counterterms are $A_{2s}$.

If one momentum vanishes the result is
\begin{align}
\CTcon 3 4 &\supset  2 \times \frac{3}{2^{11} m^5 L^6}
\frac1{L^6} \sum_{\vec p\ne 0} 
\frac{1+ \cO(p^2)}{p^6}
\ \ \Rightarrow\ \
a_2^\four \supset -4 \cK
\,,
\label{eq:res7}
\end{align}
where the initial factor of $2$ arises because there are two choices of vanishing
loop momentum.

If both momenta are nonvanishing then I find
\begin{align}
\CTcon 3 4 &\supset  \frac{3}{2^{11} m^5 L^6} \Bigg\{
\left[\frac1{L^3} \sum_{\vec p\ne 0}-\int_p\right]
\left[\frac1{L^3} \sum_{\vec q\ne 0}-\int_q\right]
\frac1{ \omega_p p^2}\frac1{\omega_q q^2}
\nonumber\\
&\qquad -\frac1{L^3} \sum_{\vec p\ne 0}\left[\frac1{L^3} \sum_{\vec q\ne 0}-\int_q\right]
\frac{m^2}{\omega_p p^4} \frac{1}{\omega_q q^2} 
-\frac1{L^3} \sum_{\vec q\ne 0}\left[\frac1{L^3} \sum_{\vec p\ne 0}-\int_p\right]
\frac1{\omega_p p^2} \frac{m^2}{\omega_q q^4}
\Bigg\}
\label{eq:SfishSint}
\\
\Rightarrow a_2^\four
&\supset  -4 \cI \cJ
\,.
\label{eq:SfishS}
\end{align}
All contributions here involve at least one sum-integral difference, a result that arises
due to the cancelation with counterterms. Thus there is no contribution to $\CTcon 3 4$ of 
the standard form involving $\Mthr$, Eq.~(\ref{eq:standard}).
This implies that, in order to be consistent with Eq.~(\ref{eq:DE324}), the double-fish
diagram, viewed as an infinite-volume
scattering diagram, must give a vanishing contribution to $\Mthr^\four$.

To see that this is indeed the case, I recall how the IR regulation and subtraction
of Ref.~\HSTH\ works for such a diagram.
Since the diagram diverges at threshold (due to the intermediate propagator), 
one must insert momenta, then perform the subtraction (in this case of $I_{0,\delta}$),
and then take the threshold limit.
Using the labeling in Fig.~\ref{fig:3part2}(a),
the momentum configuration chosen in Ref.~\HSTH\ 
is that the spectator momenta vanish ($\vec p=\vec k=0$), implying that
the momentum flowing through the intermediate propagator
also vanishes, $\vec q=\vec P-\vec p-\vec k=0$, while the ``nonspectator pair''
have nonzero momenta, $\vec a=-\vec b \ne 0$.
The CM energy flowing through the diagram is then
\begin{equation}
E = 3 m + \frac{a^2}{m} + \cO(a^4/m^3)
\,.
\end{equation}
The intermediate propagator is
\begin{equation}
\Delta(q) = \frac{i}{q^2-m^2+i\epsilon} = \frac{i}{(E-m)(E-3m+i\epsilon)}
\,,
\end{equation}
and the $i\epsilon$ can be dropped provided $a\ne 0$.
The scattering amplitudes attached to both vertices are then partially off shell,
since $q^2\ne m^2$.
I label them $\cM_{2,\off}^{\rm (2,S)}$, with the superscript indicating $2$ for second order
and $S$ for the s-channel loop.
According the definition in Ref.~\HSTH, this amplitude should be s-wave projected.
However, this is automatically satisfied here, since the amplitude depends only on 
$s=4 (m^2+a^2)\equiv s_a$ and not on the direction of $\vec a$.
The result is thus
\begin{equation}
i\cM_3^{(u,u)} \supset i\cM_{2,\off}^{\rm (2,S)}(s_a) \frac{i}{(E-m)(E-3m)} 
i\cM_{2,\off}^{\rm (2,S)}(s_a)
\,.
\label{eq:M3fish}
\end{equation}
The superscript $(u,u)$ on $\cM_3$ follows the notation of Ref.~\HSTH\
and indicates that the amplitude is unsymmetrized.

In order to obtain a finite threshold amplitude, the appropriate part of $I_{0,\delta}$
must be subtracted. This is~\HSTH
\begin{align}
iI_0^{(u,u)} &= i\cM_{2,s}^{\rm (2,S)}(s_a) i G^\infty_{s} i \cM_{2,s}^{\rm (2,S)}(s_a)\,,
\label{eq:I0fish}
\\
iG^\infty_s &= \frac{i}{2m(E-3m + i\epsilon)}
\,,
\end{align}
where again the $i\epsilon$ can be dropped.
Here the subscript $2,s$ indicates that this is a
contribution to the two-particle s-wave scattering amplitude.
Now I note that the s-channel loop amplitude is independent of the value of $q^2$,
so that, in fact, $\cM_{2,\off}^{\rm (2,S)}=\cM_{2,s}^{\rm (2,S)}$. 
Thus the difference between 
Eqs.~(\ref{eq:M3fish}) and (\ref{eq:I0fish}) can be simplified to
\begin{equation}
i\left(\cM_3^{(u,u)}-I_0^{(u,u)}\right) 
= i \cM_{2,s}^{\rm (2,S)}(s_a) \frac1{2m (E-m)}\cM_{2,s}^{\rm (2,S)}(s_a) 
 \xrightarrow{E\to3m}
 i \cM_{2,s}^{\rm (2,S)}(4m^2) \frac1{(2m)^2}\cM_{2,s}^{\rm (2,S)}(4m^2)  = 0
 \,.
\end{equation}
In the last step I have taken the threshold limit by sending $a\to 0$,
which is possible since the IR divergence has canceled. 
I find that the result vanishes in this limit because, by definition, 
all second and higher-order contributions to the scattering amplitude 
vanish at threshold.
The final step is to symmetrize the result, which does not change the
fact that the result vanishes.
Thus the double-fish diagram does not contribute to $\Mthr^\four$.

\subsubsection{Fish and sinker diagram: Fig.~\ref{fig:3part2}(b)}
\label{sec:1S1Tfish}

This diagram combines
with the $A_{2t}+A_{2u}$ part of Fig.~\ref{fig:3part2}(d), 
the $A_{2s}$ part of Fig.~\ref{fig:3part2}(e), and the $(A_{2t}+A_{2u})A_{2s}$ part of
Fig.~\ref{fig:3part2}(f), together with horizontal reflections.
When evaluating the contribution to $\CTcon 3 4$,
the momentum integrals in the counterterms can be converted into sums
at the order I work. I then find that the total summand vanishes identically,
so that there is no contribution to $\CTcon 3 4$.
As for the double-fish diagram, this implies that, if Eq.~(\ref{eq:DE324}) is to hold,
then the fish and sinker diagram must give no contribution to $\Mthr^\four$.

The argument that this is the case is more subtle than for the double-fish diagram,
because the off-shell two-to-two amplitude appearing at the right-hand vertex
in Fig.~\ref{fig:3part2}(b) now depends on $q^2$.
I label this amplitude $\cM_{2,\off}^{\rm (2,T)}(s_a,t_a,u_a)$, with the superscript
T indicating the
t/u-channel loop. Because it is off shell it depends on all three Mandelstam variables,
and thus on $q^2$ through the relation $s_a+t_a+u_a=3m^2 + q^2$.
For the kinematic configuration explained above
the Mandelstam variables are $s_a=4(m^2+a^2)$ and $t_a=u_a=m(3m-E)$.
Since these are independent of the direction of $\vec a$, the off-shell
amplitude is pure s-wave so there is no need to apply the s-wave projection.
The IR subtraction now takes the form (before symmetrization)
\begin{equation}
 i\cM_{2,\off}^{\rm (2,S)}(s_a) \frac{i}{(E-m)(E-3m)} i\cM_{2,\off}^{\rm (2,T)}(s_a,t_a,u_a)
-
i\cM_{2,s}^{\rm (2,S)}(s_a) \frac{i}{2m(E-3m)} i \cM_{2,s}^{\rm (2,T)}(s_a)\,,
\end{equation}
where in the second term $\cM_{2,s}^{(2,\rm T)}$ is the contribution to the on-shell, s-wave, 
two-particle amplitude coming from the t/u-channel diagram. I stress that
this is not the same as $\cM_{2,\off}^{\rm (2,T)}$, although the difference vanishes at threshold
since both quantities vanish there.
The difference can be rewritten as
\begin{equation}
 i\cM_{2,s}^{\rm (2,S)}(s_a) \frac{i}{2m(E-m)} i\cM_{2,s}^{\rm (2,T)}(s_a)
+
i\cM_{2,s}^{(2,S)}(s_a) \frac{i}{2m(E-3m)} 
\left[i \cM_{2,\off}^{\rm (2,T)}(s_a,t_a,u_a)- i\cM_{2,s}^{\rm (2,T)}(s_a)\right]\,,
\end{equation}
Both terms are finite when $E\to 3m$, 
but again the limiting value is zero because $\cM_{2,s}^{\rm (2,S)}$ vanishes at threshold.

\subsubsection{Double-sinker diagram: Fig.~\ref{fig:3part2}(c)}
\label{sec:1T1Tfish}

This diagram combines with the contributions from
Figs.~\ref{fig:3part2}(e) and (f) in which the counterterms are $A_{2t}+A_{2u}$.
Here both loop sums can be converted to integrals, and the cancelation with the
counterterms is exact. Thus there is no contribution to $\CTcon 3 4$.

Because of this I expect no contribution from these diagrams also to $\Mthr^\four$.
Including the subtraction, the result for the unsymmetrized amplitude is
\begin{equation}
i\cM_{2,\off}^{\rm (2,T)}(s_a,t_a,u_a) \frac{i}{(E-m)(E-3m)} 
i\cM_{2,\off}^{\rm (2,T)}(s_a,t_a,u_a)
-
i\cM_{2,s}^{\rm (2,T)}(s_a) \frac{i}{2m(E-3m)} i \cM_{2,s}^{\rm (2,T)}(s_a) \,.
\label{eq:doublesinker}
\end{equation}
Using the facts that 
$\cM_{2,\off}^{\rm (2,T)}(s_a,t_a,u_a)-\cM_{2,s}^{\rm (2,T)}(s_a) \propto (E-3m)$
[with the explicit form of this difference given in Eq.~(58) of Ref.~\HSTH]
and $\cM_{2,s}^{\rm (2,T)}(4m^2)=0$,
the difference (\ref{eq:doublesinker}) can be shown to vanish when $E\to 3m$.

\begin{figure}[tbh]
\begin{center}
\includegraphics[scale=0.5]{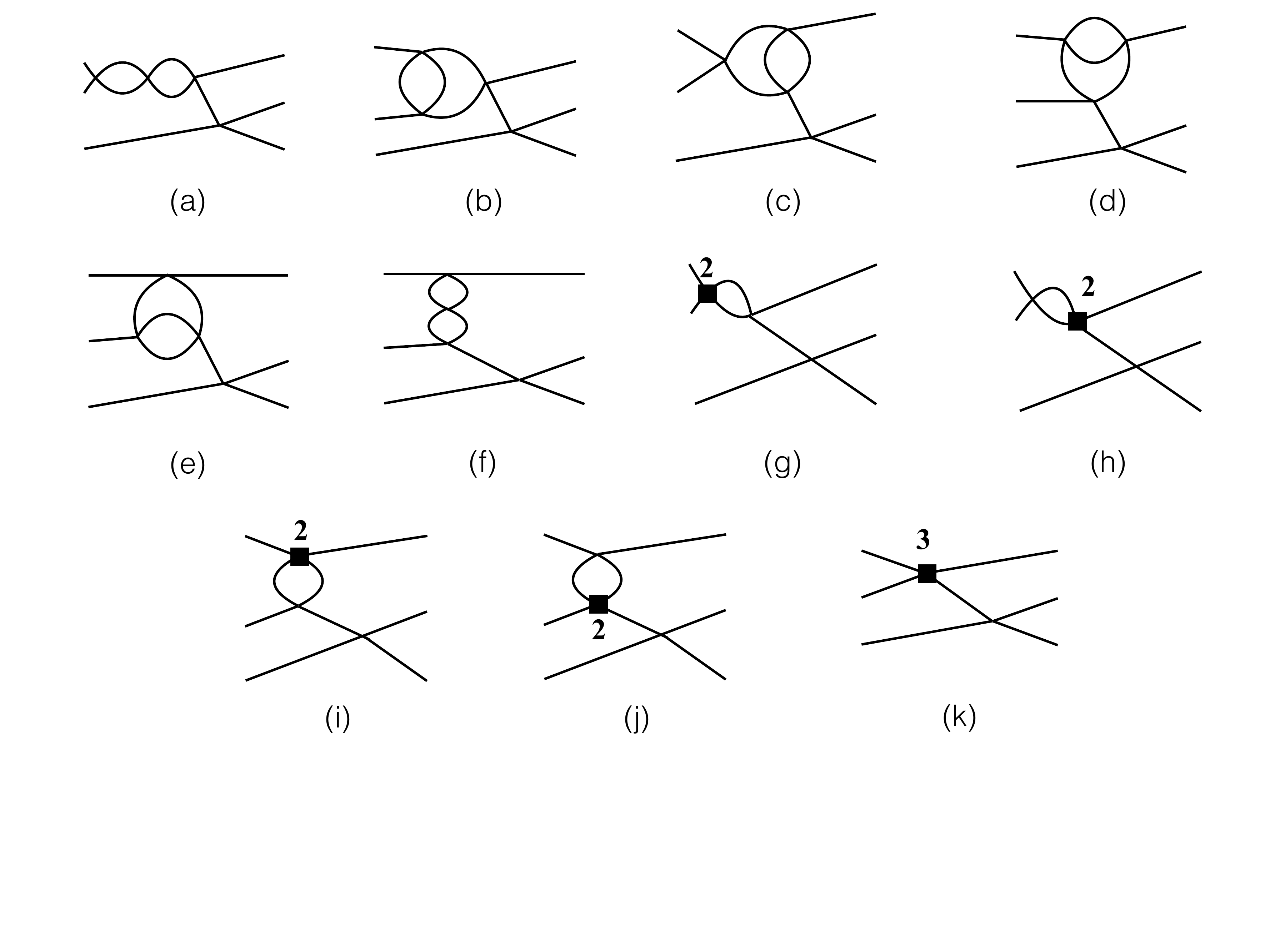}
\vskip -1.truein
\caption{Further diagrams contributing to $\CTcon 3 4$ and $\Mthr$. Reflections are not shown.}
\label{fig:3part3}
\end{center}
\end{figure}

\subsubsection{SS single-fish diagram: Fig.~\ref{fig:3part3}(a)}
\label{sec:ssfish}

I now move to radiative corrections to the 
three-particle tree diagram that involve a single two-loop correction.
These diagrams are shown in Fig.~\ref{fig:3part3},
and I begin with that involving a ``fish" with two s-channel loops,
Fig.~\ref{fig:3part3}(a), 
which combines with the $A_{2s}$ parts of 
Figs.~\ref{fig:3part3}(g) and (h) 
and the $A_{3ss}$ part of Fig.~\ref{fig:3part3}(k),
along with their horizontal reflections.

If both momenta are nonzero, the result has the same form
as that for the double-fish diagram, Eq.~(\ref{eq:SfishSint}), except for an overall
additional factor of 2 because of the reflected diagram.
Similarly there are contributions when one of the loop momenta vanishes that are
twice those from the double-fish diagram.
Thus in total I find
\begin{align}
a_2^\four &\supset 8 \cI \cJ - 8 \cK
\,.
\label{eq:SSfish}
\end{align}

Turning now to the contribution to $\Mthr^\four$,
the subtracted amplitude takes the form
\begin{equation}
 i\cM_{2,\off}^{\rm (3,SS)}(s_a) \frac{i}{(E-m)(E-3m)} (-i)
-
i\cM_{2,s}^{\rm (3,SS)}(s_a) \frac{i}{2m(E-3m)} (-i )
\,,
\label{eq:SSdiff}
\end{equation}
where the superscript indicates the third-order contribution 
involving two  s-channel loops. As above, the s-channel loops give a result
that does not depend on the off-shellness, $q^2$, so
$\cM_{2,\off}^{\rm (3,SS)} = \cM_{2,s}^{\rm (3,SS)}$.
Thus the difference (\ref{eq:SSdiff}) can be written as
\begin{equation}
i \cM_{2,s}^{\rm (3,SS)}(s_a) \frac1{m(m-E)} 
\xrightarrow{E\to 3m} 
-i \cM_{2,s}^{\rm (3,SS)}(4m^2) \frac1{2m^2}
\,,
\end{equation}
which vanishes because $\cM_{2,s}^{\rm (3,SS)}(4m^2)=0$.

\subsubsection{TS fish diagram: Fig.~\ref{fig:3part3}(b)}
\label{sec:tsfish}

This diagram combines with the $A_{2t}+A_{2u}$ part of
Fig.~\ref{fig:3part3}(g), and the $A_{3st}$ part of Fig.~\ref{fig:3part3}(k),
plus reflections. 

In this case the contribution to $\CT 3 4$ does not vanish because the
cancelation with the counterterms is not exact. The result can be written
(after converting counterterm integrals into sums) as
\begin{equation}
\CTcon 3 4 \supset  \frac{3}{2^{9} m^5 L^6}
\frac1{L^6} \sum_{\vec p, \vec q} 
f^{\rm TS}(\vec p, \vec q)
\,.
\end{equation}
The sum is convergent in the IR and UV and thus can be replaced by
an integral at the order I work.
The result of the numerical integration is
\begin{align}
\CTcon 3 4 &\supset  \frac{3}{2^{9} m^5 L^6} \frac{I^{\rm TSf}}{8\pi^4}
\,,\qquad
I^{\rm TSf}= I^{\rm STS0}\,.
\label{eq:res9}
\end{align}
The agreement with $I^{\rm STS0}$ is to the accuracy of the numerical evaluation.

Turning now to the contribution to $\Mthr$, I again find it vanishes.
The argument is as for the SS fish diagram, and relies on
the result that $\cM_{2,\off}^{(3,{\rm TS})}$ depends
only on $s_a$ (and not on $q^2$), implying that it equals $\cM_{2,s}^{(3,{\rm TS})}$,
and also that $\cM_{2,s}^{(3,{\rm TS})}(4m^2)=0$.

\subsubsection{ST fish diagram: Fig~\ref{fig:3part3}(c)}
\label{sec:stfish}

The ST fish diagram combines with the $A_{2t}+A_{2u}$ part of
Fig.~\ref{fig:3part3}(h) and the $A_{3st}$ part of Fig.~\ref{fig:3part3}(k), plus reflections.

As for the TS diagram the contribution to $\CT 3 4$ involves an IR and UV convergent sum
that can be converted to an integral. Numerical evaluation leads to
\begin{align}
\CTcon 3 4 &\supset  \frac{3}{2^{9} m^5 L^6} \frac{I^{\rm STf}}{8\pi^4}
\,,\qquad
I^{\rm STf} = 0.355066
\,.
\label{eq:ISTf}
\end{align}

The ST fish diagram also gives a nonvanishing contribution to $\Mthr^\four$.
This can be written as
\begin{equation}
i\Mthr^{\rm (4,STf)} = \lim_{E\to 3m}2 \cS \left\{
 i\left[\cM_{2,\off}^{(3,\ST)}(s_a,t_a,u_a) -\cM_{2,s}^{(3,\ST)}(s_a)\right]
 \frac{i}{q^2-m^2}
  (-i)
  \right\} \,,
\label{eq:MthrTS}
\end{equation}
where the overall factor of 2 arises from the reflection,
and $\cS$ indicates symmetrization over the choice of initial and final state
spectator particle (which in the end simply leads to a factor of $9$).
The superscripts indicate the third-order ST 
scattering diagram contained in Fig.~\ref{fig:3part3}(c),
and I have simplified using the result that $\cM_{2,s}^{(3,{\rm ST})}(4m^2)=0$.
The difference between off- and on-shell amplitudes is proportional to
$q^2-m^2$ and thus leads to a finite result in the limit.

To determine this result requires calculating the off-shell two-loop amplitude
near threshold. This is closely related to the calculation of the two-loop contribution
to the scattering length presented in Appendix~\ref{app:K}, and thus I present the
details of the off-shell calculation in Appendix~\ref{app:MthrST}.
The result is that
\begin{equation}
\Mthr^{\rm (4,STf)} = 9 \frac {1}{m^2(4\pi)^4} I^{\rm STM}
\,,\qquad
I^{\rm STM} =0.214978\,.
\label{eq:MthrST}
\end{equation}
Combining this with the result (\ref{eq:ISTf}) leads to the total contribution
from this diagram
\begin{equation}
\CTcon 3 4 \supset \frac{\Mthr^{\rm (4,STf)}}{48 m^3 L^6} +
\frac{3(I^{\rm STf}-I^{\rm STM})}{2^{12}\pi^4 m^5 L^6}
\,.
\label{eq:res10}
\end{equation}

\subsubsection{TT fish diagram: Fig.~\ref{fig:3part3}(f)}
\label{sec:ttfish}

This diagram combines with the $A_{2t}$ parts of
Figs.~\ref{fig:3part3}(i) and (j) and the $A_{3tt}+A_{3uu}$ 
part of Fig.~\ref{fig:3part3}(k), together with reflections.

The contribution to $\CT 3 4$ is simple enough to reproduce in full:
\begin{align}
\CTcon 3 4 &\supset  \frac{3}{2^{9} m^5 L^6}
\frac1{L^6} \sum_{\vec p, \vec q} 
\frac{m^3}
{4 \omega_p^3 (\omega_p+m) \omega_q^3 (\omega_q+m) (\omega_p + \omega_q)}
= \frac{3}{2^{9} m^5 L^6} \frac{I^{\rm TT}}{32 \pi^4} + \cO(L^{-7})
\,.
%I^{\rm TT} &= \frac{2(\pi-3)}{3} = 0.09440
\end{align}
The value of $I^{\rm TT}$ is given in Eq.~(\ref{eq:ITT}).

There is also a potential contribution to $\Mthr^\four$ that, before symmetrization,
has the form
\begin{equation}
2 i\left[\cM_{2,\off}^{\rm (3,TT)}(s_a,t_a,u_a) -\cM_{2,s}^{\rm (3,TT)}(s_a)\right]
 \frac{i}{q^2-m^2}
  (-i)
\,.
\label{eq:MthrTT}
\end{equation}
Since the two t-channel loops factorize, however, I can use the result
from Ref.~\HSPT\ that a single such loop gives a contribution proportional to $q^2-m^2$.
This implies that the contribution of two loops is proportional
to $(q^2-m^2)^2$, so that the overall result vanishes at threshold when $q^2 \to m^2$.

\subsubsection{SU fish diagram: Fig.~\ref{fig:3part3}(d)}
\label{sec:sufish}

This diagram combines with the $A_{2s}+A_{2u}$ part of
Fig.~\ref{fig:3part3}(i) and the $A_{3tu}/2$ 
part of Fig.~\ref{fig:3part3}(k), as well as the reflections of all diagrams.

The contribution to $\CT 3 4$ is again a finite integral
\begin{align}
\CTcon 3 4 &\supset  
%\frac{3}{2^{8} m^5 L^6}
%\frac1{L^6} \sum_{\vec p, \vec q} f^{SU}(\omega_p,\omega_{pq},\omega_q)
%= 
\frac{3}{2^{8} m^5 L^6}
\frac{I^{\rm SU}}{16\pi^4} \,,
%\qquad I^{SU} = 0.039656 \,.
\end{align}
where $I^{\rm SU}$ is given in Eq.~(\ref{eq:ISU}).

The contribution to $\Mthr^\four$ is worked out in Appendix~\ref{app:MthrSU},
yielding
\begin{align}
\Mthr^{\rm (4, SUf)} &= 9 \frac{1}{m^2(4\pi)^4} (- I^{\rm SUr})
\,,\qquad
I^{\rm SUr} = -0.274156\,.
\label{eq:MthrSUf}
\end{align}
Thus in total the SU fish diagram gives
\begin{equation}
\CTcon 3 4 \supset \frac{\Mthr^{\rm (4, SUf)}}{48 m^3 L^6} +
\frac{3(I^{\rm SU}+I^{\rm SUr})}{2^{12}\pi^4 m^5 L^6}
\,.
\label{eq:CTcon34SUf}
\end{equation}

\subsubsection{US fish diagram: Fig.~\ref{fig:3part3}(e)}
\label{sec:usfish}

This is the final diagram of this class,
and combines with the $A_{2s}+A_{2u}$ part of
Fig.~\ref{fig:3part3}(j) and the $A_{3tu}/2$ 
part of Fig.~\ref{fig:3part3}(k), as well as reflections.

The contribution to $\CT 3 4$ is 
\begin{align}
\CTcon 3 4 &\supset  
%\frac{3}{2^{8} m^5 L^6}
%\frac1{L^6} \sum_{\vec p, \vec q} f^{US}(\omega_p,\omega_{pq},\omega_q)
%= 
\frac{3}{2^{8} m^5 L^6}
\frac{I^{\rm USf}}{16\pi^4}
\,,\qquad
I^{\rm USf} =-0.13788
\,.
\label{eq:CTcon34USfa}
\end{align}
The contribution to $\Mthr^\four$ is worked out in Appendix~\ref{app:MthrUS},
and leads to
\begin{equation}
\CTcon 3 4 \supset \frac{\Mthr^{\rm (4, USf)}}{48 m^3 L^6}
+ \frac{3(I^{\rm US}-I^{\rm USM})}{2^{12}\pi^4 m^5 L^6}
\,,\qquad
I^{\rm USM} =0.096623\,.
\label{eq:CTcon34USf}
\end{equation}

\subsection{Remaining diagrams}
\label{sec:remaining}

\begin{figure}[tbh]
\begin{center}
\includegraphics[scale=0.5]{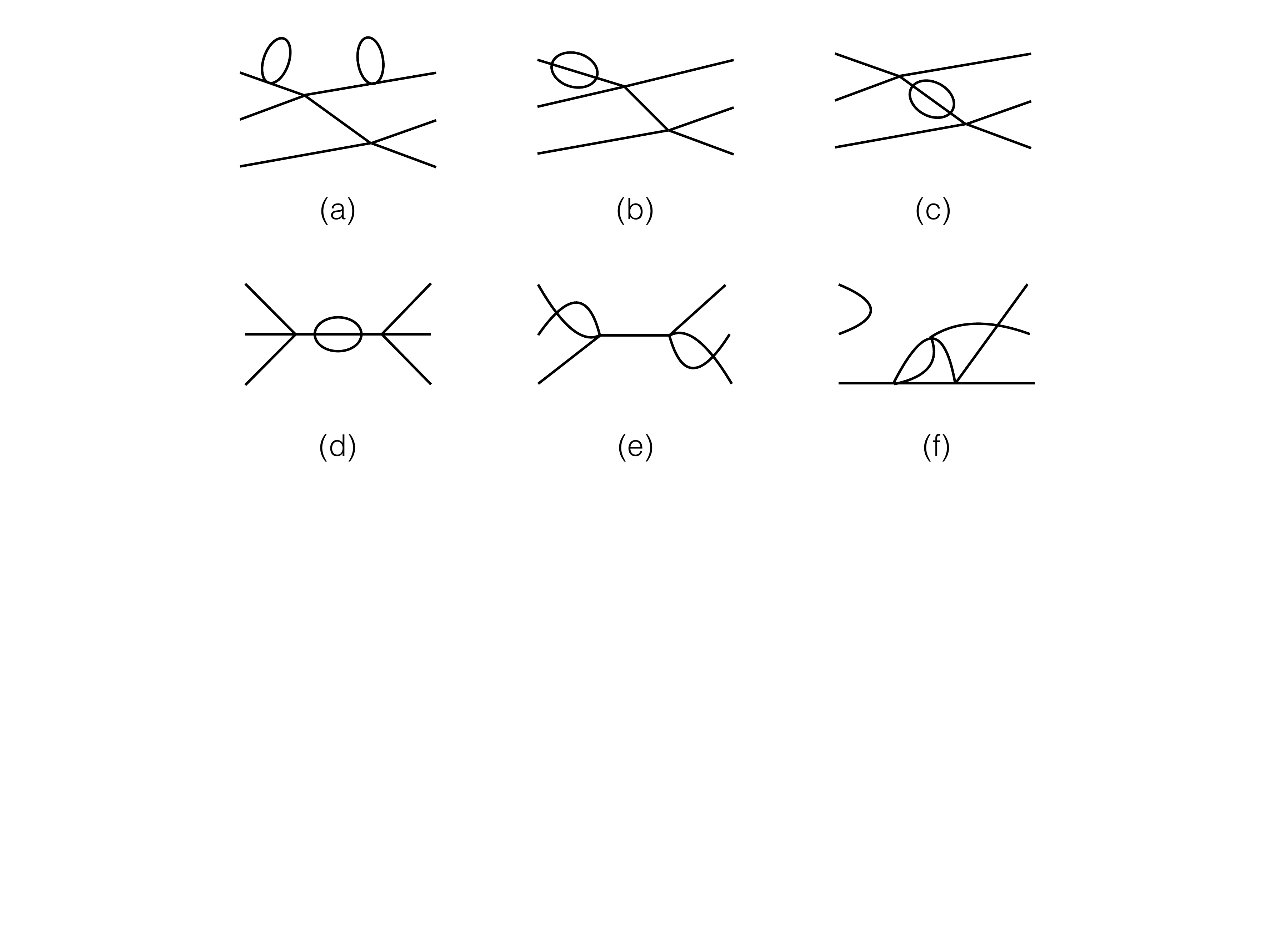}
\vskip -2.5truein
\caption{Examples of Feynman diagrams contributing to $\CTcon 3 4$
that do not lead to nonstandard finite-volume dependence.
See text for further discussion.}
\label{fig:renorm3}
\end{center}
\end{figure}

Finally, I discuss diagrams of various classes that turn out either not to contribute,
or to contribute only results of the standard form, Eq.~(\ref{eq:standard}).

The first class are those with mass and wave-function renormalization parts.
Examples of their contributions to $\CTcon 3 4$
are shown in Figs.~\ref{fig:renorm3} (a), (b) and (c).
The arguments of Sec.~\ref{sec:renorm2} can be used to show that,
when combined with the corresponding counterterms these 
either cancel completely,
which is the case for Fig.~\ref{fig:renorm3}(a),
or lead to contact terms,
as is true for Figs.~\ref{fig:renorm3}(b) and (c).
These contact terms then either lead to contributions to the amplitudes $A_{j,k}$ 
of Eq.~(\ref{eq:amps}) alone, and not to $\Delta E_{3,\thr}$,
as is the case for Fig.~\ref{fig:renorm3}(b),
or lead to contributions to $\CTcon 3 4$ of the standard form, Eq.~(\ref{eq:standard}),
an example being Fig.~\ref{fig:renorm3}(c).

There is one diagram, that of Fig.~\ref{fig:renorm3}(d), requiring special
treatment, because it has a physical cut through the renormalization part. 
This implies that there is a
nonzero remainder when the momentum sums in this part are replaced by integrals.
However, by explicit calculation I find that this remainder is subleading in $1/L$
because the IR divergence is rather weak.
Thus this diagram also gives only the standard contribution of Eq.~(\ref{eq:standard}).

There are also many other diagrams that are one-particle reducible, e.g.
Fig.~\ref{fig:renorm3}(e). Although some of these diagrams, such as this example,
have three-particle cuts, the resulting summands only have $1/p^2$ IR divergences
and so sums can be converted to integrals at the order I work. Thus all such
diagrams lead to contributions of the standard form, Eq.~(\ref{eq:standard}).

Finally, there are partially disconnected diagrams such as Fig.~\ref{fig:renorm3}(f).
As explained in Ref.~\HSPT, these amount to studying the three-particle
threshold energy using, on one or both sides, an operator that creates a single particle,
and can be dropped.

\subsection{Summary}

Combining the results from Eqs.~(\ref{eq:res1}), (\ref{eq:DTresc}),
(\ref{eq:res2}), (\ref{eq:res3}), (\ref{eq:res4}), (\ref{eq:res5}), 
(\ref{eq:res7}), (\ref{eq:SfishS}), (\ref{eq:SSfish}), (\ref{eq:res9}), (\ref{eq:res10}),
(\ref{eq:CTcon34SUf}), and (\ref{eq:CTcon34USf}),
and recalling the definitions Eqs.~(\ref{eq:D32calc}) and (\ref{eq:a24def}), 
leads to the following total contribution to $a_2^\four$ from $\CTcon 3 4$:
\begin{multline}
a_2^\four\Big|_{\CTcon 3 4} =
-\frac{2^{14} m^2 \pi^6}{9} \Mthr^{(4)}
+  \left[4 \cI \cJ + 24 \cK+ c_L \log(N_{\rm cut}) + \cC_4 + \cC_5 + \cC_F \right]
\\
-  2^{6} \pi^2 \left(I^{\rm STS0}+I^{\rm STf}-I^{\rm STM} +\tfrac14 I^{\rm TT}
+ I^{\rm SU}+I^{\rm SUr}+I^{\rm US}- I^{\rm USM}\right)
\,.
\label{eq:CTcon34res}
\end{multline}
Combining this result with those in Eqs.~(\ref{eq:a34CT24}) and (\ref{eq:a34CZ33})
gives the final result for $a_2^\four$:
\begin{align}
a_2^\four&=
- \frac{2^{14} m^2 \pi^6}{9} \Mthr^{(4)}
+ \left[2 \cI^2  -8 \cI\cJ +18 \cK + c_L \log(N_{\rm cut}) + \cC_4 + \cC_5 + \cC_F \right]
+ 2^6 \pi^2 {\cal R}\,,
\label{eq:a24final}
\\
{\cal R} &= 
\left[I^{\rm SST0}\!-\! I^{\rm STS0}\right] 
+ \left[I^{\rm STS0}\!-\!I^{\rm STS}\!-\!I^{\rm STf}\!+\!I^{\rm STM}\right] 
+ \left[ I^{\rm SU}\!-\! 2 I^{\rm SSU} \!-\! I^{\rm SUr}\right]
+ \left[2 I^{\rm SU} \!-\!I^{\rm US}\!-\! 2 I^{\rm SSU} \!+\! I^{\rm USM}\right]
\,.
\label{eq:a24residue}
\end{align}
This result should be compared to that obtained from the general FV formalism,
which, using Eq.~(\ref{eq:DE324}), is given by the first two terms in Eq.~(\ref{eq:a24final}).
Thus agreement requires the residue $\mathcal R$ to vanish.
In fact, each of the four quantities in square brackets
vanishes separately to within the numerical accuracy
of integration (roughly five significant figures).
This completes the desired check.
The fact that cancelations occur in subsets of quantities
indicates that the cancelations can be understood at a diagrammatic level.

\section{Conclusions}
\label{sec:conc}

The calculation presented here has confirmed the threshold expansion derived
from the three-particle quantization condition of Refs.~\cite{Hansen:2014eka,Hansen:2015zga}.
It provides a further nontrivial test of the quantization condition
as well as of the rather involved determination of the threshold expansion
from the quantization condition~\cite{Hansen:2016fzj}.

Comparing the form of the predictions for 
$\Delta E_{2,\thr}^{(4)}$ and $\Delta E_{3,\thr}^{(4)}$,
given in Eqs.~(\ref{eq:DE24}) and (\ref{eq:DE34}) respectively,
one sees that the latter contains additional ``geometric" constants,
$\cC_F$, $\cC_4$, and $\cC_5$, as well as the $\Mthr$ term.
The new constants arise from the need to subtract IR divergences from
diagrams in which there is alternating pairwise two-particle scattering.
The calculation presented here checks that these subtractions do the
job for which they were designed.
By working at fourth order all terms contributing to the IR subtraction
have been tested.

As noted in Ref.~\HSPT, the result from the present relativistic calculation cannot
be compared to those obtained using nonrelativistic quantum mechanics
 both because relativistic effects
enter at $\cO(L^{-6})$ and because the nonrelativistic analogs of $\Mthr$ differ,
in general, by finite amounts. Nevertheless, I observe that the coefficients 
of $\cI^3$, $\cI \cJ$ and $\cK$ do agree with those of Ref.~\Beane.

It will be interesting and useful to test other approaches to the three-particle quantization
condition, such as that using a nonrelativistic, particle-dimer approach~\cite{Hammer:2017uqm,Hammer:2017kms},
by confirming that they reproduce the three-particle threshold expansion.

\section*{Acknowledgments}
This work was supported in part by the United States Department of Energy grant No.
DE-SC0011637.
I thank Ra\'ul Brice\~no, Max Hansen and John Spencer for discussions and comments.

\newpage
\appendix
\section{Sum-integral differences}
\label{app:sumint}

The following two results from Ref.~\HSPT\ are used repeatedly in the main text
\begin{align}
\left[\frac1{L^3} \sum_{\vec p\ne 0} - \int_p\right] \frac{f(p^2)}{p^2} 
&= \frac{\cI f(0)}{(2\pi)^2 L} - \frac{ f'(0)}{ L^3} 
\label{eq:Idef}
\\
\frac1{L^3} \sum_{\vec p\ne 0} \frac{g(p^2)}{p^2} &= 
\frac{L \cJ g(0)}{(2\pi)^4} +
\int_p \frac{g(p^2)-g(0)}{p^4}   + \cO(L^{-1})
\,.
\label{eq:Jdef}
\end{align}
$\cI$ and $\cJ$ are numerical constants whose explicit values are not needed in this paper.
The shorthand for integrals is defined by
\begin{equation}
\int_p \equiv \int \frac{d^3p}{(2\pi)^3}\,.
\end{equation}

I also need a third result,
\begin{equation}
\frac{1}{L^3} \sum_{\vec p\ne 0} \frac{h(p^2)}{p^6}
= \frac{L^3 \cK h(0)}{(2\pi)^6} + \cO(L)
\,,
\label{eq:Kdef}
\end{equation}
where $\cK= \sum_{\vec n=0} 1/n^6$.
This equation can be derived by writing $h(p^2)=h(0) + [h(p^2)-h(0)]$
and using Eq.~(\ref{eq:Jdef}).
These results hold if $f$, $g$ and $h$ are smooth functions that in 
particular are regular at $p^2=0$. 
All three equations have, in addition to the power-law corrections shown,
exponentially suppressed corrections of the form $e^{-\sqrt{\bar m^2} L}$.
The scale $\bar m^2$ depends on the form of $f(p^2)$, but is no larger than
the position of the nearest singularity of $f$ in the complex $p^2$ plane.
For the functions I encounter this scale is usually given by the scalar mass,
$\bar m \sim m$.
The equations also assume that the sums and integrals have been regularized in
the UV, if needed, although the form of the $L$-dependent terms does not depend
on the details of the regularization.

\section{Counterterm for the coupling constant}
\label{app:counter}

The counterterm for the coupling constant has the expansion\footnote{%
The definition of $A_3$ used here differs from that of Ref.~\cite{Hansen:2015zta}.}
\begin{equation}
\delta Z_\lambda = \lambda^2 A_2 + \lambda^3 A_3 + \lambda^4 A_4 + \dots
\,.
\end{equation}
$A_2$ is identical to the quantity of the same name used in
Ref.~\cite{Hansen:2015zta}:
\begin{align}
A_2 &= A_{2s} + A_{2t}+A_{2u}\,,
\label{eq:A2def}
\\
A_{2s} &= \int_{p}^\Lambda \frac1{8 \omega_p \vec p^2}\,, \qquad
A_{2t} = A_{2u}=\int_{p}^\Lambda \frac1{8 \omega_p^3}\,.
\label{eq:A2stu}
\end{align}
The superscript $\Lambda$ indicates the necessity of UV regulation.
For most of our calculations the form of the regulator can be
left implicit, since in the end the integrals to be evaluated are UV convergent.
However, when calculating the two-loop K matrix the explicit
forms of the counterterms are needed. Using dimensional regularization, these are
\begin{align}
A_{2t} &= A_{2u} = \int \frac{ d^{d-1} q}{(2\pi)^{d-1}} \frac1{8 \omega_q^3}
= \frac{\Gamma(2\!-\!d/2)}{(4\pi)^{d/2}} \frac12\,,
\label{eq:A2tDR}
\\
A_{2s} &= \int \frac{ d^{d-1} q}{(2\pi)^{d-1}} \frac1{8 q^2 \omega_q}
= \frac{\Gamma(2\!-\!d/2)}{(4\pi)^{d/2} }\frac12 \frac1{d-3}\,,
\label{eq:A2sDR}
\end{align}
where $d$ is the number of space-time dimensions.
I stress that these are the same counterterms as obtained by 
doing the $d$-dimensional momentum integrals directly, rather than the hybrid
approach time/three-momentum approach used here.

The calculations described in the main text also require the explicit form
of $A_3$.
This is obtained by calculating the scattering amplitude at two-loop order,
including all counterterm contributions, and setting its value at threshold to zero.
A convenient way of obtaining the integrand of the scattering amplitude is
simply to calculate the diagrams contributing to $\partial_\tau C_{2,\thr}^{(3)}$,
using the methods described in Sec.~\ref{sec:methods}, 
and pick out the coefficient multiplying $1/(8 m^2 L^3)$.
The four two-loop diagrams are those in Fig.~\ref{fig:C2lam3} (d), (e), (f) and (g),
which I label, respectively, as SS, TT, SU and ST diagrams.
These are combined with the counterterm diagrams in the same way as in the calculation
of $\CT 2 3$ in Sec.~\ref{sec:DE24}.
The corresponding two-loop counterterms are given by:
\begin{align}
A_3 &= A_{3ss}+A_{3tt} + A_{3su} +A_{3st}\,,
\\
A_{3ss} &= A_{2s}^2 \,,
\\
A_{3tt} &= 2 A_{2t}^2\,,
\,,
\\
A_{3su}&= \iint_{p,q}^\Lambda \left\{
4 A_{2t} (A_{2s}+A_{2u}) - 
\left[
\frac{W_{pq}^3 + W_{pq}^2 \omega_p - W_{pq} m^2 + \omega_{pq} m^2}
{4 \omega_p^3 \omega_{pq} \omega_q (W_{pq}^2-m^2)^2}
\right] \right\}
\,,
\\
A_{3st}&= \iint_{p,q}^\Lambda \frac1{32\omega_p^3\omega_q q^2}
\left[1-
\frac{2 \omega_p^2 (W_{pq}+\omega_q)} {\omega_{pq} (W_{pq}^2-m^2)}\right]
\,,
\end{align}
where $W_{pq}=\omega_p+\omega_{pq}+\omega_q$.
%In the Feynman diagrams below, the counterterms are shown by filled boxes,
%with the associated number indicating the power of $\lambda$.

The calculations of Sec.~\ref{sec:CT24} also require parts of the counterterm $A_4$, 
specifically those needed for Figs.~\ref{fig:C2lam4}(f)-(k).
These are given by enforcing the renormalization condition, which implies that
the coefficient of $1/L^3$ in the contribution to $\CT 2 4$ of a given three-loop diagram,
together with the counterterm diagrams, must vanish.
The contributions to $A_4$
 are thus easily determined in the course of the calculation. In general their form
is long and uninformative, and I do not quote the results here.

\section{Two-loop K matrix and related quantities}
\label{app:K}

In this appendix I calculate the derivative of the
s-wave K matrix at threshold, $\cK'_{2,s,\thr}$, defined in Eq.~(\ref{eq:Kprime}),
at cubic order in $\lambda$.
This is needed in order to check that the perturbative result obtained here,
Eq.~(\ref{eq:a24fin}), agrees with the prediction from the quantization condition,
Eq.~(\ref{eq:DE24}).
In addition I determine the partially off-shell K matrix needed in order
to define $\Mthr^{(4)}$, itself needed to test the result Eq.~(\ref{eq:DE324}).
These calculations are done in infinite volume using standard momentum-space
methods for Feynman diagrams, and I only provide a sketch of the details.

\subsection{Calculation of $\cK'^{(3)}_{2,s,\thr}$}

I focus first on the on-shell K matrix, which is a real, analytic function of $q^2$,
with $q$ the momentum of each particle in the CM frame. 
It is most straightforward to use Feynman rules with the $i\epsilon$ prescription to
determine the s-wave scattering amplitude $\cM_{2,s}$, 
and then obtain $\cK_{2,s}$ using the general relation\footnote{%
This expression holds above threshold. In order that $\cK_{2,s}$ be a real, analytic
function of $q^2$, one must replace $iq$ with $-|q|$ below threshold.}
\begin{equation}
\cK_{2,s}(q^2) = \cM_{2,s}(q) - \cM_{2,s}(q) \frac{i q}{16\pi E} \cM_{2,s}(q)
+  \cM_{2,s}(q) \frac{i q}{16\pi E} \cM_{2,s}(q) \frac{i q}{16\pi E} \cM_{2,s}(q)
+ \cO(\lambda^4)\,.
\label{eq:K2fromM2}
\end{equation}
Here $E$ is the total CM energy.
Expanding $\cK_{2,s}$ and $\cM_{2,s}$ in powers of $\lambda$,
e.g. $\cK_{2,s}=\sum_{n=1}^\infty \lambda^n \cK_{2,s}^{(n)} $,
and noting that there is no $q$ dependence at leading order,
I have $\cM_{2,s}^{(1)}=\cK_{2,s}^{(1)}=-\lambda$ for all $q$.
In our renormalization scheme I also have $\cM_{2,s}^{(n)}(0)=\cK_{2,s}^{(n)}(0)= 0$
for $n\ge 2$.
Since $\cK_{2,s}$ has no term linear in $q$, it follows from Eq.~(\ref{eq:K2fromM2}) that
\begin{equation}
\cM_{2,s}^{(2)} = \frac{i q}{16\pi E} + \cO(q^2)
\,.
\end{equation}
Substituting this back into Eq.~(\ref{eq:K2fromM2}) then yields
\begin{equation}
\cK_{2,s}^{(3)} = \cM_{2,s}^{(3)} - \frac{q^2}{(16\pi E)^2} + \cO(q^3)\,.
\label{eq:K2fromM2b}
\end{equation}
It follows that the imaginary part of $\cM_{2,s}^{(3)}$ begins at $\cO(q^3)$
and that
\begin{equation}
\frac{d\cK_{2,s}^{(3)}}{d q^2}\bigg|_\thr 
= \frac{d\cM_{2,s}^{(3)}}{dq^2}\bigg|_\thr  - \frac{1}{(32\pi m)^2}\,.
\label{eq:K2fromM2c}
\end{equation}
The difference between the derivatives of $\cK^{(3)}_{2,s}$ and $\cM^{(3)}_{2,s}$ arises
only from diagrams having two physical cuts.
There is in fact only one such diagram---the SS diagram to be discussed shortly.
For all other diagrams the 
threshold derivatives of $\cK^{(3)}_{2,s}$ and $\cM^{(3)}_{2,s}$ are the same,
and so I can calculate the latter.

The diagrams that contribute to $\cM_{2,s}^{(3)}$ are those of Fig.~\ref{fig:C2lam3},
for which I use the same names as in the calculation of $\CZ 2 3$ in Sec.~\ref{sec:CZ23}.
Each two-loop diagram is combined with the same counterterm diagrams as
described in Sec.~\ref{sec:CZ23}.

\subsubsection{SS diagram, Fig.~\ref{fig:C2lam3}(d)}

This diagram factorizes into two single s-channel loops, along with corresponding
counterterms. Each of these begins at $\cO(q)$ with the imaginary term resulting from
the cut, leading to
\begin{equation}
\cM_{2,s}^{(3,SS)} = - \left(\frac{i q}{16\pi E}\right)^2 + \cO(q^3)\,.
\end{equation}
This contribution exactly cancels that appearing in Eq.~(\ref{eq:K2fromM2b}), so that
\begin{equation}
\cK_{2,s}^{(3,SS)} = \cO(q^3)\,.
\end{equation}
Thus this diagram gives no contribution to $d\cK_{2,s}/dq^2$ at threshold.\footnote{%
One can also understand this result by
noting that the $iq$ terms arise from the use of the $i\epsilon$ pole prescription and
are absent when using the principal value prescription appropriate for $\cK_{2,s}$.}

\subsubsection{TT diagram, Fig.~\ref{fig:C2lam3}(e)}

The result again factorizes, but in this case the contribution of each loop is
proportional to $q^2$ because there is no imaginary part
(and because the loop plus counterterm vanishes at threshold). 
Thus the product of the two loops is proportional to $q^4$ and
gives no contribution to the desired derivative.

\subsubsection{SU diagram, Fig.~\ref{fig:C2lam3}(f)}
\label{app:KSU}

Figure~\ref{fig:C2lam3}(f) combines with 
the $A_{3su}$ contribution to Fig.~\ref{fig:C2lam3}(a),
together with the $A_{2s}+A_{2u}$ contribution to Fig.~\ref{fig:C2lam3}(c),
and their vertical reflections.
In fact, the contribution proportional to $A_{3su}$ is independent of $q^2$,
and so can be ignored here.

To evaluate the diagram I follow the method described in Ref.~\PS.
I use the momentum labels shown in Fig.~\ref{fig:C2lam3}(f).
Including the vertical reflection and the contraction with $q_3$ and $q_4$
interchanged, I find
\begin{equation}
i\cM_{2,\bare}^{(3,{\rm SU})}=
2  (-i)^3
\int_{p,k} \Delta(p+q_1)\Delta(p+q_4)\Delta(p+k)\Delta(k)
\,,
\label{eq:M2s3su}
\end{equation}
where $\Delta(p)= i/(p^2-m^2 + i\epsilon)$, and ``bare"
indicates that the counterterm has not yet been included. Throughout this appendix,
and in contradistinction to the main text and the other appendices, I use 
$p,\, k, \dots$ to refer to four-momenta, and the
shorthand $\int_p = \int d^dp/(2\pi)^d$ for the dimensionally regulated
four-momentum integral.

From Eq.~(\ref{eq:M2s3su}) it is clear that the only Lorentz invariant combinations of
external momenta that can appear are $q_1^2=m^2$,  $q_4^2=m^2$ and
$(q_1 -q_{4})^2 = t$.
Dependence on $q^2$ enters only through
$t = - 2 q^2 (1 - \cos \theta_{14})$,
and the derivative with respect to $q^2$ at threshold thus picks out the term linear in $t$.
Since in this term the $s$-wave projection averages $\cos\theta_{14}$ to zero,
it follows that
\begin{equation}
\frac{d \cK_{2,s}^{(3, {\rm SU})}}{dq^2}\bigg|_\thr
= - 2 \frac{d \cM_{2}^{(3,{\rm SU})}}{dt}\bigg|_\thr
\,.
\label{eq:derivrelation}
\end{equation}
This equality also holds for the counterterm diagram Fig.~\ref{fig:C2lam3}(c),
since it also depends only on $t$.

Introducing Feynman parameters $x$ and $y$ for the $k$ and $p$ integrals,
respectively, performing the $k$ integral, and combining denominators using
Eq.~(10.56) of Ref.~\PS, I reach\footnote{%
For brevity, in the following I set the scale introduced by dimensional
regularization, $\mu$, equal to unity. The final answer does not depend on $\mu$.}
\begin{equation}
i\cM_{2,\bare}^{(3,{\rm SU})}=
-2 \int_y \int_x \int_w
(1-w) w^{1-d/2}
\frac{\Gamma(4-d/2)}{(4\pi)^{d/2}}
\int_p\frac{1}{D_{{\rm SU}}^{4-d/2}}
\,,
\end{equation}
where the factor in the denominator is
\begin{align}
D_{{\rm SU}} &= 
w\left[m^2 - x(1-x) p^2\right]
+ (1-w) \left[m^2 -p^2 - y (2 p\cdot q_{1}+q_1^2) - (1-y) (2 p\cdot q_4+q_4^2)\right]
\,,
\end{align}
and the $y$, $x$ and $w$ integrals range from $0$ to $1$.
Expanding out $D_{st}$ and completing the square
by shifting from $p$ to $p'$ (whose explicit form is not needed) leads to
\begin{align}
D_{{\rm SU}} &= - \gamma p'^2 + m^2 + \Delta_{st}\,,
\\
\gamma &= 1-w + w x (1-x)\,,
\\
\Delta_{{\rm SU}}(w) &= 
\frac{1-w}{\gamma} \left\{ - m^2 w x (1-x) - t (1-w) y (1-y) 
- (q_4^2-m^2) w x (1-x)(1-y)\right\}
\,.
\label{eq:DeltaSU}
\end{align}
Here I have set $q_1^2=m^2$, but shown the $q_4^2$ dependence
for later use in the discussion of $\Mthr$.
In the rest of this subsection I will set $q_4^2=m^2$.
Note that, although $\Delta_{\rm SU}$ depends on $w$, $x$, $y$ and $t$, it is convenient
to show only its dependence on $w$ explicitly.
Performing the $p$ integral leads to the result
\begin{equation}
\cM_{2,\bare}^{(3,{\rm SU})} =
- 2 
\frac{\Gamma(4-d)}{(4\pi)^{d}}
\int_y\int_x\int_w
\frac{(1-w) w^{1-d/2}}{\gamma^{d/2}}
\frac1{(m^2+\Delta_{{\rm SU}}(w))^{4-d}}
\,.
\label{eq:su_after_p}
\end{equation}
The denominator $m^2+\Delta_{{\rm SU}}(w)$ does not vanish in the physical
region ($t \le 0$). 
%The nearest nonanalyticity occurs at $t>4 m_\pi^2$, when the
%denominator vanishes for some choices of $x$, $y$, $w$.

The result (\ref{eq:su_after_p}) has an explicit UV pole from the $\Gamma(4-d)$,
and the integral over $w$ near 0 leads to a second pole. 
Lying underneath the double pole is a momentum-dependent single pole.
This is  canceled by the contribution from the counterterm diagram
Fig.~\ref{fig:C2lam3}(c), which yields
\begin{equation}
\cM_{2,\ct}^{(3,{\rm SU})} =
 2 
\frac{\Gamma(2-d/2)^2}{(4\pi)^{d} (m^2)^{2-d/2}}  F_{{\rm SU}}
\int_0^1 dy
\frac1{(m^2+\Delta_{{\rm SU}}(0))^{2-d/2}}
\,.
\label{eq:su_counterterm}
\end{equation}
The factor of 
\begin{equation}
F_{{\rm SU}}=\frac{1 + 1/(d-3)}{2} % = \frac{d-2}{d-3} 
= 1 - \frac{d-4}{2} + \dots
\end{equation}
arises from the counterterm $A_{2s}+ A_{2u}$ 
[see Eqs.~(\ref{eq:A2tDR}) and (\ref{eq:A2sDR})].

Combining Eqs.~(\ref{eq:su_after_p}) and (\ref{eq:su_counterterm})
in a way that allows the double pole to be made explicit gives
\begin{align}
\cM_{2}^{(3,{\rm SU})} &=
- 2 
\frac{\Gamma(4-d)}{(4\pi)^{d}}
\int_y \int_x\int_w\, w^{1-d/2}
\left[\frac{(1-w)}{\gamma^{d/2}}
\frac1{(m^2+\Delta_{{\rm SU}}(w))^{4-d}}
- 
\frac1{(m^2+\Delta_{{\rm SU}}(0))^{4-d}} \right]
\nonumber\\
&\quad-  2 
\frac{1}{(4\pi)^{d}} 
\int_y \left[
\frac{\Gamma(4-d)}{2-d/2}
\frac1{(m^2+\Delta_{{\rm SU}}(0))^{4-d}} 
- \frac{\Gamma(2-d/2)^2 F_{{\rm SU}}}{(m^2)^{2-d/2}}
\frac1{(m^2+\Delta_{{\rm SU}}(0))^{2-d/2}}\right]
\,.
\label{eq:su2}
\end{align}
Strictly speaking this is not the full renormalized result for the amplitude:
the first integral leads to a momentum-independent single pole that will be canceled by
the $A_{3su}$ counterterm, while the second leads to momentum-independent
double and single poles that will similarly be canceled.
However, these poles are irrelevant for the momentum dependence of interest here.

Evaluating the derivative with respect to $t$ and setting $t=0$, I find an expression
in which one can set $d=4$:
\begin{align}
\frac{d}{d t}\cM_{2,s}^{(3,{\rm SU})}\bigg|_\thr &=
- 2 
\frac1{(4\pi)^4 m^2} \left\{
\int_y \int_x\int_w\, \frac{y(1-y)}{w}
\left[\frac{(1-w)^3}{\gamma^{2}[\gamma-w(1-w)x(1-x)]}-1\right]
- \frac16 \right\}
\,.
\label{eq:su_deriv2}
\end{align}
The expression in curly braces has the numerical value
$I^{\rm SUr} =-0.274156$.
Converting to the desired derivative using Eq.~(\ref{eq:derivrelation})
leads to the final result:
\begin{equation}
\cK'^{(3,{\rm SU})}_{2,s} \equiv
m^2 \frac{d \cK_{2,s}^{(3,{\rm SU})}}{dq^2}\bigg|_\thr 
%= 4 \lambda^3 \frac1{(4\pi)^4} \frac{(-0.644934-1)}{6}
=
\frac{I^{\rm SUr}}{2^6 \pi^4}
\,.
\label{eq:KSU}
\end{equation}
%Here the label ``r" is a reminder that this integral arises from a calculation of the scattering length.

I have checked this result in two ways.
The first is to calculate $\cM_{2,s}^{(3,{\rm SU})}$ using a
finite-volume correlation function in which the external momenta are nonvanishing.
The second involves relating the SU result to that from the ST diagram by crossing and
then calculating the desired derivative for the ST diagram using
an unsubtracted dispersion relation. This in turn requires the imaginary part of the
amplitude from the ST diagram, which can be obtained from the results
in the following subsection.

\subsubsection{ST diagram, Fig.~\ref{fig:C2lam3}(g)}
\label{app:KST}

The calculation proceeds as for the SU diagram, except for the following changes.
First, the number of Wick contractions differs, leading to a reduction by an overall
factor of 2.
Second, the counterterm contribution here is proportional to $A_{2t}+A_{2u}$, so
$F_{\rm SU}$ is replaced by $F_{\rm ST}=1$.
Finally, the crossing transformation needed to go from the SU to the ST diagram
necessitates the substitutions $q_4\to q_3$ and $q_1\to -q_4$.
This implies that $t=(q_1-q_4)^2$ is replaced by $(q_3+q_4)^2=s$, 
so $\cM_2^{(3,{\rm ST})}$ is pure s-wave.
Since $s=4(q^2+m^2)$ the desired derivative is given by
\begin{equation}
\frac{d \cK_{2,s}^{(3,{\rm ST})}}{dq^2}\bigg|_\thr
= 4 \frac{d \cM_{2}^{(3,{\rm ST})}}{ds}\bigg|_\thr
\,,
\label{eq:derivrelationb}
\end{equation}
where threshold occurs at $s=4m^2$.
Since the ST diagram has a physical cut,
$\cM_{2,s}^{(3,{\rm ST})}$ has an imaginary part, but, as explained
above, this begins only at $\cO(q^3)$, and does not contribute to the derivative at threshold.

Thus I find that $\cM_{2,s}^{(3,{\rm ST})}$ is given by Eq.~(\ref{eq:su2})
except that the overall factor of $2$ is dropped, $F_{SU}\to 1$,
and $\Delta_{SU}(w)$ is replaced by
\begin{equation}
\Delta_{{\rm ST}}(w) =
\frac{1-w}{\gamma} \left\{ - m^2 w x (1-x) - s (1-w) y (1-y) - (q_3^2-m^2) w x (1-x)(1-y)\right\}
\,,
\end{equation}
Here I have set $q_4^2=m^2$ but kept the dependence on $q_3^2$ explicit
for use in the calculation of $\Mthr$. For the rest of this subsection, however,
I set $q_3^2=m^2$.

The calculation in this case is more challenging than for the SU diagram because
individual terms  lead to a diverging  derivative at threshold that becomes finite
only when they are combined. Thus I proceed somewhat differently, expanding
about $d=4$ and keeping only the $s$-dependent part, which is finite and has the form
\begin{align}
\cM_{2,s}^{(3,{\rm ST})} &\supset
\frac1{(4\pi)^{4}} \left\{
\iiint_0^1 dy \,dx\, dw\, \frac1w
\left[\frac{(1-w)}{\gamma^{2}} \log D_{{\rm ST}}(w)
- \log D_{{\rm ST}}(0) \right]
- \frac12
\int_0^1 dy \left[\log D_{{\rm ST}}(0)\right]^2 \right\}
\,,
\label{eq:st}
\end{align}
where 
\begin{equation}
D_{{\rm ST}}(w) = 1 + \Delta_{{\rm ST}}(w)/m^2\,.
\end{equation}
As long as $s < 4m^2$ all terms are real, and the $y$ integrals can be done analytically.
Doing the remaining $w$ and $x$ integrals numerically, and then determining the
$s$ derivative also numerically, I find
\begin{equation}
\cK'^{(3,{\rm ST})}_{2,s,\thr} = 4 m^2 \frac{d\cM_{2,s}^{(3,{\rm ST})}}{d s}\bigg|_\thr =
\frac{I^{\rm STr}}{(4\pi)^4 }
\,,\qquad
I^{\rm STr}= 1.14009
\,.
\label{eq:KST}
\end{equation}
I have checked this result using an unsubtracted dispersion relation for 
$d \cM_{2,s}^{(3,{\rm ST})}/ds$.

\subsection{Calculation of $\cM_{2,\off}^{(3)}$}

The off-shell two-loop two-particle scattering amplitude is needed in
the determination of $\Mthr$
in Secs.~\ref{sec:stfish}, \ref{sec:sufish} and \ref{sec:usfish}.
The calculation of this amplitude is closely related to that of the
effective range from the ST and SU diagrams described above.

\subsubsection{ST diagram}
\label{app:MthrST}

As explained in Sec.~\ref{sec:stfish}, what is needed is
the amplitude $\cM_{2,\off}^{\rm (3,{\rm ST})}$, which arises from Fig.~\ref{fig:C2lam3}(g)
when $q_3$ is off shell while the remaining external legs are on shell.
The form of the result has already been described in Sec.~\ref{app:KST}: it is
given by Eq.~(\ref{eq:su2}) except that $F_\SU$ is set to unity, 
and $\Delta_\SU$ is replaced by
\begin{equation}
\Delta_{\ST,\off}(w) =
\frac{1-w}{\gamma} \left\{ - m^2 w x (1-x) - s (1-w) y (1-y) - \delta w x (1-x)(1-y)\right\}
\,.
\end{equation}
Here I have parametrized the off-shellness of $q_3$ by
\begin{equation}
\delta = q_3^2-m^2\,.
\end{equation}
There is one difference from Sec.~\ref{app:KST}, which concerns the overall factor.
In Eq.~(\ref{eq:su2}) this is $2$, while in Sec.~\ref{app:KST} it is $1$.
Here it changes to $1/2$, because the horizontal reflection (the TS diagram) does not
contribute to $\Mthr$, as explained in Sec.~\ref{sec:tsfish}.

The specific quantity of interest is the difference between off- and on-shell amplitudes
defined in Eq.~(\ref{eq:MthrTS}), which 
[noting that $q$ in Eq.~(\ref{eq:MthrTS}) is called $q_3$ here]
is given by %\footnote{%
%%
%Strictly speaking setting $s=4m^2$ automatically sets $\delta=0$, but I write the
%latter equation explicitly for the sake of clarity.}
%%
\begin{equation}
\Mthr^{\rm (4,{\rm STf})} = 
2 \times 9 \times 
\frac{d \cM_{2,\off}^{\rm (3,{\rm ST})}}{d\delta}\bigg|_{s=4m^2,\delta=0}
\,.
\label{eq:Mthr4STf}
\end{equation}
The factor of $2$ arises from the horizontal reflection, and the $9$ 
results from symmetrization.
Noting that $\Delta_{\ST,\off}(0)$ is independent of $\delta$, I find 
\begin{align}
\frac{d \cM_{2,\off}^{\rm (3,\ST)}}{d\delta}\bigg|_{s=4m^2,\delta=0}
&=
\frac12 \frac{1}{(4\pi)^{4}}
\int_y \int_x \int_w
\frac{(1-w) }{\gamma^{2} w}
\left\{ \frac1{m^2 + \Delta_{\ST,\off}(w)}
\frac{d \Delta_{\ST,\off}(w)}{d\delta} \right\}\bigg|_{s=4m^2,\delta=0}
\,,
\\
&\equiv
\frac12 \frac{1}{(4\pi)^{4}m^2} I^{\rm STM}\,,
\qquad I^{\rm STM}=-0.214978
\,.
\end{align}
In the first line I have set $d=4$ since the result is finite,
and in the second performed the numerical evaluation of the integral.
Substituting this result in Eq.~(\ref{eq:Mthr4STf}) leads to
the result quoted in the main text, Eq.~(\ref{eq:MthrST}).

\subsubsection{SU diagram}
\label{app:MthrSU}

Here the calculation turns out to be identical to that of Sec.~\ref{app:KSU},
aside from overall factors.
This is because the off-shellness enters only through $t$,
so $\Delta_\SU(w)$ retains the same form as in the on-shell case.
The overall factor is reduced by 2
because here SU and US diagrams must be considered separately.
Thus I find
\begin{equation}
\cM_{2,\off}^{\rm (3,\SU)}(t) = \frac12 \cM_{2,s}^{\rm (3,\SU)}(t)
\,.
\label{eq:M2SUoff}
\end{equation}
The contribution to $\Mthr$ is
\begin{align}
\Mthr^{\rm (4,{\rm SUf})} &= 18
\frac{d \cM_{2,\off}^{\rm (3,\SU)}}{d \delta}\bigg|_{t=0,\delta=0}
\\
&= \frac{9}2 \frac{d \cM_{2,s}^{\rm (3,\SU)}}{d t}\bigg|_{t=0}
\,.
\label{eq:MthrSU}
\end{align}
where in the second step I have used
\begin{equation}
t = \tfrac12 \delta - \tfrac12 (s-4m^2)
\,,
\label{eq:tSU}
\end{equation}
as well as Eq.~(\ref{eq:M2SUoff}).
The required derivative is given in Eq.~(\ref{eq:su_deriv2}),
\begin{equation}
\frac{d \cM_{2,s}^{\rm (3,\SU)}}{d t}\bigg|_{t=0}
= - \frac{2 I^{\rm SUr}}{(4\pi)^4 m^2}
\,.
\label{eq:ISUr2}
\end{equation}
Inserting this into Eq.~(\ref{eq:MthrSU}) gives the result quoted
in the main text, Eq.~(\ref{eq:MthrSUf}).

\subsubsection{US diagram}
\label{app:MthrUS}

The contribution of the US diagram to $\Mthr$ differs from that for the SU fish diagram 
because, when Fig.~\ref{fig:C2lam3}(g) is vertically reflected, $q_4$ and $q_3$ are interchanged.
Thus $\Delta_\SU$ is replaced by
\begin{equation}
\Delta_{\US,\off}(w) = 
\frac{1-w}{\gamma} \left\{ - m^2 w x (1-x) - t (1-w) y (1-y) 
- \delta w x (1-x)(1-y)\right\}
\,,
\end{equation}
with $t$ still given by Eq.~(\ref{eq:tSU}).
The contribution to $\Mthr$ is thus 
\begin{align}
\Mthr^{\rm (4,USf)} &= 18
\frac{d \cM_{2,\off}^{\rm (3,US)}}{d \delta}\bigg|_{t=0,\delta=0}
\\
&= \frac92 \frac{d \cM_{2,s}^{\rm (3,SU)}}{d t}\bigg|_{t=0}
- 
\frac{9}{(4\pi)^4 m^2}
\left\{\int_y\int_x\int_w
\frac{(1-w^2)}{\gamma^{2}} 
\frac{ 2 x (1-x)(1-y) }
{\gamma - w(1-w) x (1-x)}\right\}
\,,
\label{eq:MthrUSb}
\\
&\equiv \frac{9}{(4\pi)^4 m^2} I^{USM}
\,,\qquad
I^{\rm USM}=0.096623\,.
\label{eq:MthrUS}
\end{align}
To obtain the final result I have used Eq.~(\ref{eq:ISUr2}) and
the fact that the triple integral in Eq.~(\ref{eq:MthrUSb}) 
evaluates to $-0.177533$.
Combining this result with Eq.~(\ref{eq:CTcon34USfa}) leads
to the result quoted in the main text, Eq.~(\ref{eq:CTcon34USf}).

\bibliographystyle{apsrev4-1} %%% physical review
\bibliography{ref} %%% ref.bib file
	
\end{document}